%% file: iccv_review.tex
\documentclass[10pt,twocolumn,letterpaper]{article}
\pdfoutput=1

\usepackage{iccv}
\usepackage{times}
\usepackage{epsfig}
\usepackage{graphicx}
\usepackage{amsmath}
\usepackage{amssymb}
\usepackage{subcaption}
\usepackage{float}
\usepackage{amsthm}

\theoremstyle{definition}
\newtheorem{definition}{Definition}[section]

\usepackage[breaklinks=true,bookmarks=false]{hyperref}

\iccvfinalcopy 

\def\iccvPaperID{4042} 

\ificcvfinal\fi

\begin{document}

\title{Photon-Starved Scene Inference using Single Photon Cameras}

\author{Bhavya Goyal \quad \quad Mohit Gupta\\
University of Wisconsin-Madison\\
{\tt\small \{bhavya, mohitg\}@cs.wisc.edu}\\
}
\maketitle
\ificcvfinal\thispagestyle{empty}\fi

\input{00-abstract.tex}
\input{01-introduction.tex}

\input{02-related.tex}
\input{03-imagingmodel.tex}

\input{04-photonscalespace.tex}

\input{05-experiments.tex}

\input{06-monodepthestimation}

\input{07-realexperiments.tex}

\input{08-conclusion.tex}

{
\small
\bibliographystyle{ieee_fullname}
\bibliography{egbib}
}

\onecolumn
\section{Supplementary Report for "Photon-Starved Scene Inference using Single Photon Cameras"}
\input{09-techreport}

\end{document}

%% file: 00-abstract.tex
\begin{abstract}
Scene understanding under low-light conditions is a challenging problem. This is due to the small number of photons captured by the camera and the resulting low signal-to-noise ratio (SNR). Single-photon cameras (SPCs) are an emerging sensing modality that are capable of capturing images with high sensitivity. Despite having minimal read-noise, images captured by SPCs in photon-starved conditions still suffer from strong shot noise, preventing reliable scene inference. We propose photon scale-space -- a collection of high-SNR images spanning a wide range of photons-per-pixel (PPP) levels (but same scene content) as guides to train inference model on low photon flux images. We develop training techniques that push images with different illumination levels closer to each other in feature representation space. The key idea is that having a spectrum of different brightness levels during training enables effective guidance, and increases robustness to shot noise even in extreme noise cases. Based on the proposed approach, we demonstrate, via simulations and real experiments with a SPAD camera, high-performance on various inference tasks such as image classification and monocular depth estimation under ultra low-light, down to $<1$ PPP. Project Page: \url{https://wisionlab.cs.wisc.edu/project/photon-net}
\end{abstract}

%% file: 01-introduction.tex
\section{Scene Inference in Low Light}
Over the past decade, deep learning has achieved unmatched accuracy on several complex, real-world scene inference tasks. As these techniques have matured, a new axis in the performance space is emerging, driven by applications (e.g., autonomous navigation), where reliable performance under non-ideal imaging conditions is as important as the overall accuracy. In such safety-critical applications, it is important to consider the \emph{worst case} performance of the vision system to ensure robust \emph{all-weather operation}. For example, for a vision system to be deployed on an autonomous car, it must perform reliably across the entire range of imaging scenarios, including night-time and poorly-lit scenes, and high-speed moving objects, all of which result in photon-starved images. Even state-of-the-art inference algorithms tend to fail for such images where the sensor has simply not collected sufficient light. 

\begin{figure*}
\begin{center}
\includegraphics[width=\linewidth]{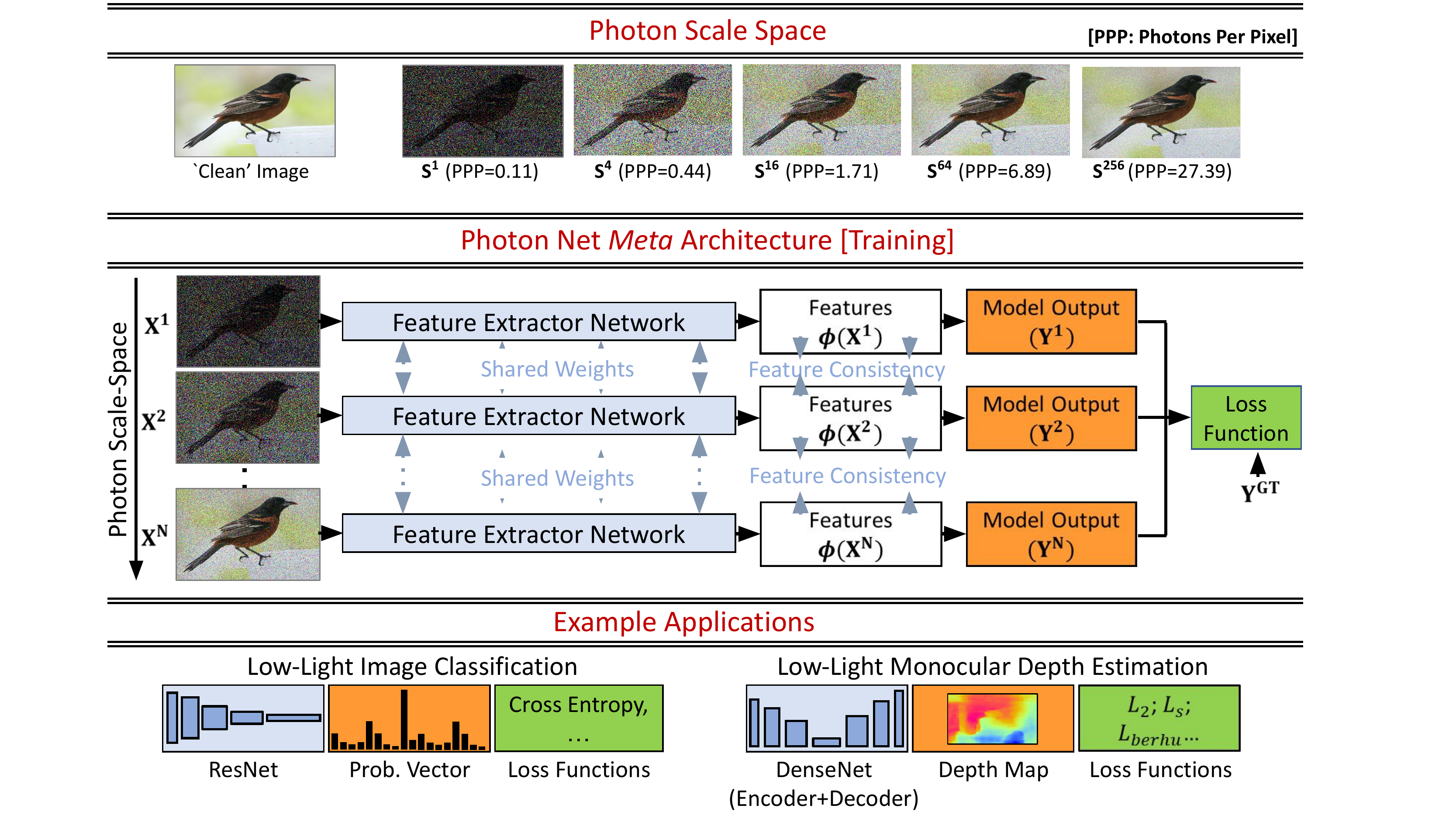}
\end{center}
\vspace{-15pt}
\caption{{\bf Inference in Low-Light using Photon Scale-Space.} {\bf (Top)} Photon scale-space is a hierarchy of images, each with a different flux level, but sharing the same scene content. Successive images in the hierarchy have similar flux, so that high-flux images can guide the low-flux images during a training procedure. {\bf (Middle)} We use photon-scale space to develop a meta network architecture called the photon net, where a network is trained with multiple input images with the same scene content but with varying noise levels
in order to push them together in the feature space. {\bf (Bottom)} The proposed approach is modular and versatile, lending itself to a wide range of inference tasks such as classification and depth estimation.}
\label{fig:photonnetarch}
\vspace{-5pt}
\end{figure*}


The goal of this paper is to develop vision systems that achieve high accuracy even in ultra low-light, when a camera pixel may receive even less than one photon per pixel. In such extreme conditions, images captured by conventional cameras get overwhelmed by noise, causing the signal-to-noise ratio (SNR) to dip below the threshold required for downstream inference algorithms to extract meaningful scene information. We propose a two-pronged approach to achieve these goals: (a) Leverage a class of highly sensitive single-photon detectors, and (b) Develop inference algorithms that are optimized for low-flux operation. \smallskip

\noindent {\bf Single-Photon Sensors:} Single-photon avalanche diodes (SPADs)~\cite{niclass2005design,rochas2003single} are an emerging image sensor technology that are capable of detecting individual incident photons with high timing precision. In the past, these sensors were limited to single pixel or low-resolution devices (e.g., 32x32 pixels), and thus restricted to scientific applications~\cite{buttafava2015non,o2018confocal,bruschini2019single}. But, recently, due to their compatibility with CMOS fabrication processes, high resolution cameras (up to 1 MPixel) have been developed based on SPADs~\cite{morimoto2020megapixel}, as well as the jots~\cite{ma2017photon} technology. These single photon cameras are capable of capturing sequences of binary frames with minimal read noise~\cite{ulku2018512}, thus opening the possibility of capturing high-quality images even in low-flux conditions. \smallskip

\noindent {\bf High-level Inference on Low-Flux Images:} So far, SPADs have primarily been used for recovering image intensities~\cite{antolovic_dynamic_2018,Ingle:2019,ma2020quanta} and low/mid-level scene information such as 3D shape~\cite{oconnor,RENKER200648,Dautet:93,Kirmani58,Shin_2016_naturecomm,Gupta:2019,Gupta:2019b} and motion~\cite{gyongy2018single}. Can we go beyond low-level imaging and signal processing, and develop algorithms for direct, high-level inference from SPAD cameras? Despite low read noise, the stochastic nature of photon arrival results in considerable shot noise in single-photon binary frames captured by SPAD cameras. Although there has been some recent work on joint denoising and classification~\cite{liu2020connecting,liu2019transferable,diamond2017dirty}, inference on ultra-low-light images where each pixel receives less than a photon on average still remains an intractable problem. 

To address this problem, we design inference techniques based on the notion of \emph{guided training}, where a high-quality image is used as a guide for training low-quality images. This is similar in spirit to the classical guided filtering~\cite{He:2010} where a guidance image is used for low-level image processing tasks, such as denoising~\cite{He:2010} and super-resolution~\cite{Lutio:2019}. More recently, the idea of guided training has been explored in the context of student-teacher training~\cite{gnanasambandam2020image} where a teacher network pre-trained on high-quality images guides a student network operating on low-flux images. These approaches rely on underlying similarities in the inputs of the student and teacher networks to aid the guidance process, and therefore, are not very effective in the extreme case where the student's and teacher's input images may have a huge difference in the number of photons-per-pixel (e.g., $<1$ vs. $>1000$). These images may have no structural similarity despite representing the same scene. If the guide and the \emph{guidee} images have no common content and features, how can one perform guided training? \smallskip

\noindent {\bf Photon-Scale Space:} We propose using a hierarchy of guide images from a wide spectrum of photon levels, each having the same scene content, but varying number of average photons-per-pixel (PPP), from as low as PPP $\approx 0.1$, going up to PPP$>100$. The key idea is that although all the images taken together span a large range of SNR values (including high SNR images at the top which provide most accurate labels), \emph{successive} images in the hierarchy have similar number of photons (and thus, similar features) so that guidance percolates down effectively to the lowest levels, to the images with the minimum PPP. We call this hierarchy of images the \emph{photon scale-space} (Fig.~\ref{fig:photonnetarch}), reminiscent of the classical image-size scale-space~\cite{Lindeberg:1994} which is used in many computer vision algorithms. \smallskip 

\noindent {\bf Photon-Net Guided Training:} Based on photon scale-space, we propose \emph{Photon Net}, a \emph{meta} architecture and training techniques for performing inference on low-flux input images (Fig.~\ref{fig:photonnetarch}). The key idea is to 
train a given network architecture with different images from a photon scale-space, so that the images having the same scene content (but different flux level) are trained together leading to effective guidance from the highest SNR training images to the low SNR test images. 
We do this by enforcing feature consistency for high-level features (e.g., the final feature vector) of the network.
Since frames at different levels in the photon scale space share the same scene content, we encourage similarity of high-level features, despite having large differences in the low-level image statistics (low/mid-level features).

We perform empirical analysis on various design choices for creating the photon scale-space (e.g., the number of levels), and suggest rules-of-thumb for good performance. Due to the known forward model of the single photon imaging process (Poisson sampling), the photon scale-space can be created using images captured from conventional cameras, making the proposed approach amenable to training using existing large-scale image datasets. We demonstrate, via extensive simulations as well as real experiments on a 1/8 megapixel SPAD array (SwissSPAD2~\cite{ulku2018512}), considerable (up to $10\%$) improvement for various inference tasks in extreme low-light conditions ($\sim$0.1 PPP). \smallskip

\noindent {\bf Scope and Implications:} The proposed approach is modular and versatile --- it is possible to use a wide range of network architectures, loss functions, and model outputs in the same framework --- thus lending itself to a variety of inference tasks including low-light image classification and even regression tasks such as monocular depth estimation in the dark (Fig.~\ref{fig:photonnetarch}). SPADs remain a nascent imaging modality, and cannot yet directly compete with conventional sensors which have been optimized over decades. However, given their sensitivity, high speed and dynamic range~\cite{antolovic_dynamic_2018,Ingle:2019,ma2020quanta}, they have the potential to provide capabilities (e.g., vision in ultra low-light and rapid motion) that were hitherto considered impossible. This work takes the first steps towards exploring SPADs as all-purpose sensors capable of not just low-level imaging, but also high-level inference across a wide gamut of challenging imaging conditions.

%% file: 02-related.tex
\section{Related Work}
\noindent {\bf Single-Photon (Quanta) Sensors:} SPADs and jots are two current major technologies for large single-photon camera arrays. Jots amplify the single-photon signal by using an active pixel with high conversion gain~\cite{fossum2005sub}. By avoiding avalanche, jots achieve smaller pixel pitch, higher quantum efficiency and lower dark current, but have lower temporal resolution~\cite{ma2017photon}. Although we demonstrate our approaches using SPADs, the computational techniques are applicable generally to a single-photon sensors, including jots.\smallskip

\noindent {\bf Inference on Single Photon Sensors:} Starting with the early (primarily theoretical) work~\cite{chen2016vision,chen2017seeing} which proposed the idea of directly performing computer vision tasks on stream of photons instead of forming an image, there has been a growing trend of using quanta sensors for various scene inference applications. This includes high-speed tracking using quanta sensors~\cite{gyongy2018single}, and more recently, object identification~\cite{antsiperov2019target} and image classification~\cite{gnanasambandam2020image}. Our work is a next step in this direction, providing a general and versatile approach capable of achieving high performance across a wide variety of scene inference tasks. \smallskip

\noindent {\bf Low-light Classification:} There has been a lot of work on inference in low-light using conventional cameras as well. The most notable in this line of work are recent approaches that perform joint denoising and inference on noisy images~\cite{liu2020connecting,liu2019transferable,diamond2017dirty}. Although such joint denoising and inference techniques outperform conventional sequential denoising and inference approaches, they do not have the benefit of effective guidance from high SNR images, and thus are unable to achieve high performance in extreme low-light conditions ($\sim 0.1$ PPP).  \smallskip

\noindent {\bf Image-size Scale Space:} A recent work~\cite{xu20203d} proposes techniques that use image-size scale space, i.e., images at multiple resolutions, for designing pose estimation techniques that can perform well for very low resolution images. We borrow numerous insights from this work, as we create photon scale-space and photon net family of architectures for inference on very low-light images. 


%% file: 03-imagingmodel.tex
\section{Passive Single Photon Imaging Model}\label{sec:imageFormation}
For a single photon camera, the number $Z(x,y)$ of photons arriving at pixel $(x,y)$ during an exposure time of $\tau$ seconds is modeled as a Poisson random variable~\cite{6104150}, whose distribution is given as:
\vspace{-5pt}
\begin{equation}
P\{Z=k\} = \frac{ (\phi \tau \eta)^k  e^{-\phi \tau \eta}}{k!} \,\,,
\end{equation}

\noindent where $\phi(x,y)$ is the photon flux (photons/seconds) incident at $(x,y)$, and $0 \leq \eta \leq 1$ is the quantum efficiency of the pixels. In the binary mode, each pixel detects at most one photon during the exposure time and returns a binary value $B(x,y)$ such
that $B(x,y) = 1$ if $Z(x,y) \geq 1$; $B(x, y) = 0$ otherwise.\footnote{After each photon detection, a SPAD pixel experiences a dead time during which it cannot detect any further photons~\cite{rochas_spad_phd_2003}. For modern SPAD pixels, the dead time is significantly smaller than the exposure time $\tau$, and therefore is not considered in the following analysis.} Due to the randomness in photon arrival, the binary measurements $B(x,y)$ are also random variables with Bernoulli distribution:
\vspace{-5pt}
\begin{equation}
\begin{aligned}
P\{B = 0\} = e^{-(\phi\tau\eta + r_q \tau)}, \\
P\{B = 1\} = 1 - e^{-(\phi\tau\eta + r_q \tau)}
\end{aligned}
\label{eq:imagingmodel}
\end{equation}
where $r_q$ is the dark count rate (DCR), which is the rate of spurious photon detections.\smallskip

\noindent {\bf Sources of Image Noise:} Conventional sensors measure incident photons as analog current, which is then converted to a discrete number. This analog-to-digital conversion (ADC) results in a fixed read noise per frame, which leads to low signal to noise ratio (SNR) in dark scenes. In contrast, SPCs directly measure the photon counts, skipping the intermediate ADC, thereby avoiding read noise. 


Although SPCs have minimal read noise, binary frames still have extremely low signal-to-noise ratio (SNR) in low flux environments due to shot noise. Fig.~\ref{fig:photonnetarch} shows an example of a clean image, with the corresponding binary image ($\mathcal{S}^1$). The shot noise in the binary image (Eq.~\ref{eq:imagingmodel}) causes extreme degradation. While it is possible to increase the SNR by temporally averaging a large number of binary frames, this approach is not applicable in the presence of scene / camera motion due to motion blur or large computational requirements of motion compensation algorithms~\cite{chi_eccv_2020,ma2020quanta}. This raises the following question: Is it possible to extract meaningful scene information from a single (or a small number of) single-photon binary frames? 



%% file: 04-photonscalespace.tex

\section{Photon Scale Space}
To address this question, we develop a \emph{guided training} approach, where high SNR images act as a guide for training low SNR images. To facilitate such guided training, we propose the concept of \emph{photon scale-space}, a hierarchy of guide images with varying flux levels, but each having the same scene content. The key idea is that although all the images taken together span a large range of SNR values (including high SNR and most informative images at the top), \emph{successive} images in the hierarchy have similar SNR levels (and thus, similar features) so that guidance percolates down effectively to the lowest levels. \smallskip

\noindent {\bf How to generate a photon scale space?} Consider a `clean image' as captured by a camera in high-flux conditions. Assuming the pixel intensities in the clean image to be the ground-truth flux values for the corresponding scene points\footnote{In general, the pixel intensities have a non-linear relationship to incident flux due to sensor's radiometric response and image compression algorithms. Although we do not explicitly model these effects, they can be accounted for in the following discussion.}, we can generate multiple stochastic binary images as captured by a single photon camera using the image formation model described in Section~\ref{sec:imageFormation}. Assuming the scene is stationary, i.e. there is no motion between binary frames, we can simulate a series of images with different flux levels by summing a sequence of $N$ binary frames (for various values of $N$) to get N-sum images ($\mathcal{S}^N$), defined as follows:

\begin{definition}[N-Sum image $\mathcal{S}^N$]
The average of N binary frames
\vspace{-10pt}
\begin{equation}
\begin{aligned}
\mathcal{S}^{N}(x,y) = \sum_{i=1}^N B_i(x,y)   \,\,.
\end{aligned}
\end{equation}
\end{definition}

Using the definition of N-Sum images, we define Photon Scale Space as a hierarchy of images with successively higher flux levels as follows:

\begin{definition}[Photon Scale Space $PSS(K, L, n)$]
A set of n N-Sum images, starting from the lowest-SNR image $S^{K}$ (noisiest), to the highest-SNR image $S^{L}$, with $K < L.$
\end{definition}

We choose the parameters $(K, L, n)$ so that the images span a large gamut of SNR levels (i.e., $K \ll L$). The choice of the number of levels $n$ presents a tradeoff: To ensure effective guidance from high SNR to low SNR images, the successive images in the hierarchy should have similar flux levels, thus requiring a large $n$. On the other hand, a large $n$ would increase computational cost of the training algorithms. In our implementations, we choose images with $N$ increasing as a geometric series $N \in [K,K(L/K)^{\frac{1}{n-1}},K(L/K)^{\frac{2}{n-1}}...,L]$ so that the approximate ratio of the flux level between successive images is a constant. We round the values of $N$ to the nearest integer if it is a fraction. 

For instance, suppose we want to train inference models for $\mathcal{S}^1$ images (1 binary frame), but use high flux images up to $\mathcal{S}^{256}$ (256 binary frames) for guidance during training. The photon scale space for this setting with, say, $5$ levels would consist of $\mathcal{S}^1, \mathcal{S}^4, \mathcal{S}^{16}, \mathcal{S}^{64}$ and $\mathcal{S}^{256}$ images.
Fig.~\ref{fig:photonnetarch} shows an example of images from photon scale space with $K=1$, $L=256$ and $n=5$, thereby spanning a broad range of SNR levels, while ensuring that successive images have similar SNR and features. \smallskip



\noindent{\bf What is the range of flux values spanned by a photon scale space?} Since each binary frame is independent, the expected value of the sum image $\mathcal{S}^N(x,y)$ is:
\begin{equation}
\begin{aligned}
E[\mathcal{S}^{N}(x,y)] = N*E[B(x,y)]\\
= N (1 - e^{-(\phi\tau\eta + r_q \tau)}) \,.
\end{aligned}
\end{equation}

The maximum likelihood estimate (MLE) of the incident flux ($\phi$) is therefore given as
\begin{equation}
\begin{aligned}
\hat{\phi}(x,y) = -\ln(1 - \mathcal{S}(x, y)/N )/\tau\eta  -  r_q/\eta \,.
\end{aligned}
\end{equation}
This non-linear relationship between $\mathcal{S}$ (number of photons detected by the camera), and $\phi$ (number of photons incident of camera) has an asymptotic behavior~\cite{ma2020quanta,antolovic_dynamic_2018} --- $\mathcal{S}^{N}$ keeps increasing with increasing number of incident photons $\phi$, allowing us to span a large range of incident flux levels in the photon scale space, even with a finite range of $N$ values.

\section{Guided Training with Photon Scale Space}

In this section, we design a guided training technique that leverages photon scale space images for developing high-performance low-light inference algorithms. \smallskip

\noindent {\bf Photon Net:} The key enabling component of the proposed technique is a \emph{meta} architecture called Photon Net that uses photon scale space images as input, along with a feature consistency loss that encourages similar feature representations for all the images belonging to the same photon scale space (thus having the same scene content), despite having a large variation in brightness levels. 

Fig.~\ref{fig:photonnetarch} shows the overview of the architecture, which consists of several identical network branches. During training, each branch takes as input an image from the photon scale space with a certain PPP level (ranging from low SNR to high SNR images). All the branches are trained with shared weights, so that gradient updates from high PPP branches can guide low PPP branches. In order for high SNR images to act as a guide to low SNR images, all the photon scale space images with different PPP levels from the same original image are trained together by sampling them in the same mini-batch. Since, the weights are shared, there is no additional overhead of network parameters as we do not keep multiple copies of the network. \smallskip

\noindent {\bf Encouraging Feature Consistency:} In order to encourage consistency in the learned feature representations for different inputs of the network (images with the same scene content but different noise levels), we use feature consistency loss during training.
It is possible to use a variety of loss functions such as contrastive loss, L2, or L1 loss for consistency. In our implementation, we used an L2 loss function (Mean Squared Error loss), to push features from the same image with different PPP levels closer to each other.
\begin{equation}
\begin{aligned}
L_{MSE}(\{x_i\}) =  \frac{1}{N} \sum_{i,j} \|\varphi(x_i) - \varphi(x_j)\|_2^2
\end{aligned}
\label{eq:contrastiveloss}
\end{equation}
where $\{(x_i)\}$ is set of all training images, $N$ is the total number of training pairs in the mini-batch with same scene content (i.e. $x_i$ and $x_j$ are images from the same original image with different PPP level) and $\varphi(.)$ is the feature output from network. We use feature vector from the final layer of the CNN (after global pooling layer) as our feature representation.

Overall loss function is the combination of $L_{MSE}$ and the primary loss function for the inference task. For the case of image classification: $L_{overall} = L_{CE} + \lambda L_{MSE}$, where $L_{CE}$ is cross entropy loss and $\lambda$ is the weighting factor to control the contribution of both losses. Please see the supplementary report for details.

\begin{figure}
    \centering
    \includegraphics[width=\linewidth]{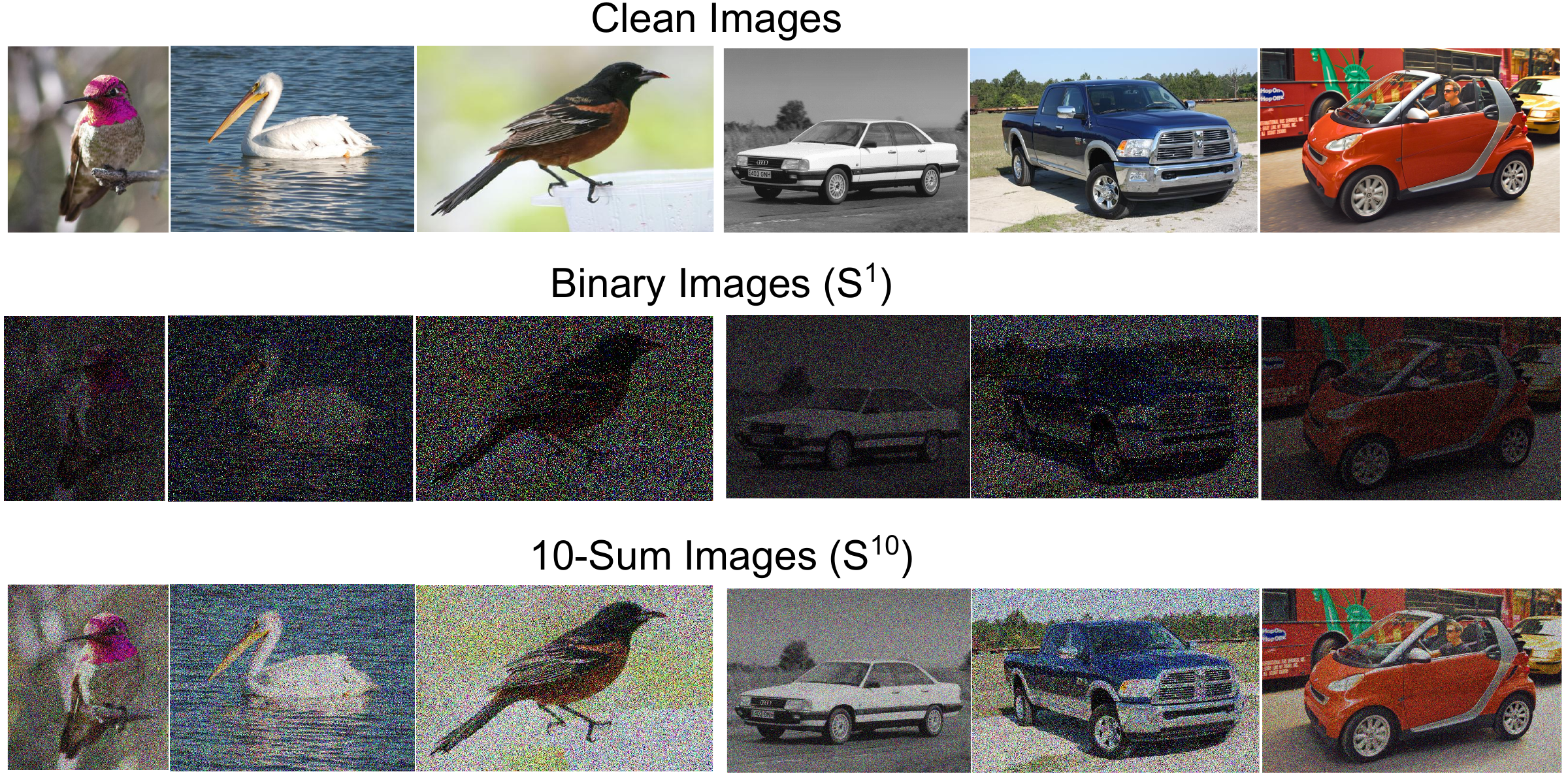}
\caption{\textbf{Simulated Single-Photon Images}: Clean images and simulated noisy images from CUB-200-2011 and CARS-196 dataset. SPCs captures sequence of binary images like ($\mathcal{S}^1$) with heavy shot noise. 10-sum images ($\mathcal{S}^{10}$) are average of 10 binary images.}
\label{fig:samplesimulatedimages}
\vspace{-0.1in}\end{figure}

%% file: 05-experiments.tex
\section{Applications: Low-Light Scene Inference}\label{sec:experiments}
The guided training approach based on photon scale space and photon net is modular since it is possible to use a wide range of network architectures, loss functions, and tasks in the same framework (Fig.~\ref{fig:photonnetarch}). We demonstrate the effectiveness and versatility of the proposed techniques via two low-flux inference tasks: image classification and monocular depth estimation.

\begin{figure}
\begin{subfigure}{\linewidth}
    \centering
    \includegraphics[width=0.8\linewidth]{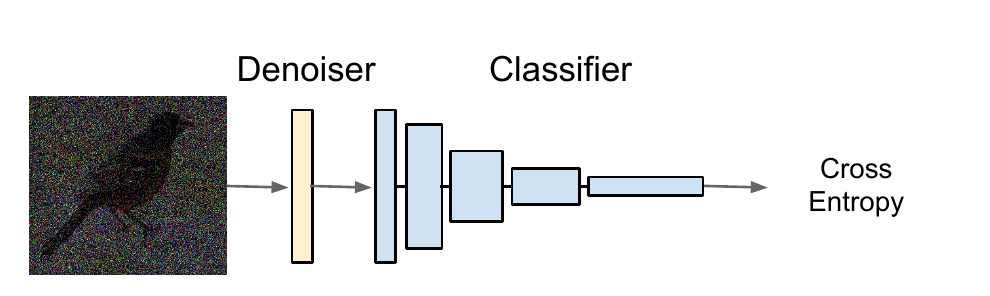}
    \vspace{-7pt}
    \caption{Joint Denoising}
    \label{fig:dirtypixelarch}
\end{subfigure}
\begin{subfigure}{\linewidth}
    \centering
    \vspace{2pt}
    \includegraphics[width=0.8\linewidth]{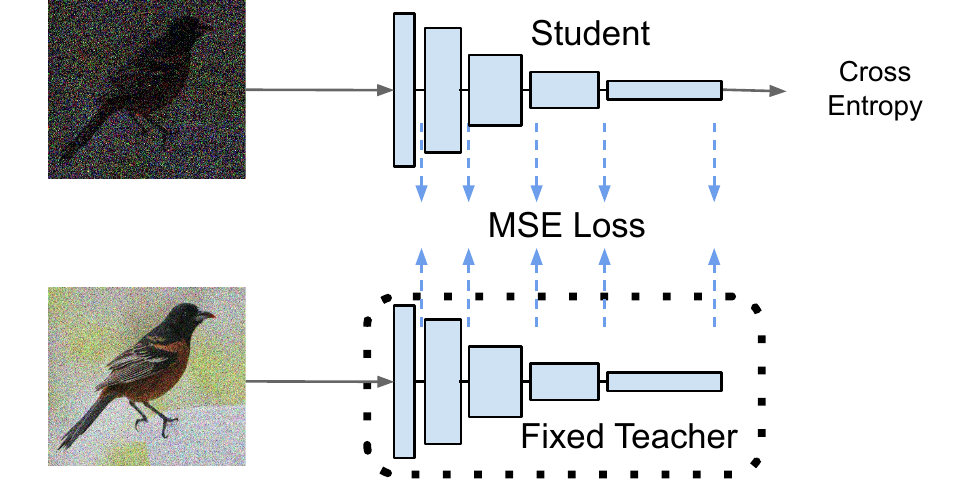}
    \vspace{-2pt}
    \caption{Student Teacher Learning}
    \label{fig:perceptarch}
\end{subfigure}
\begin{subfigure}{\linewidth}
    \centering
    \includegraphics[width=0.8\linewidth]{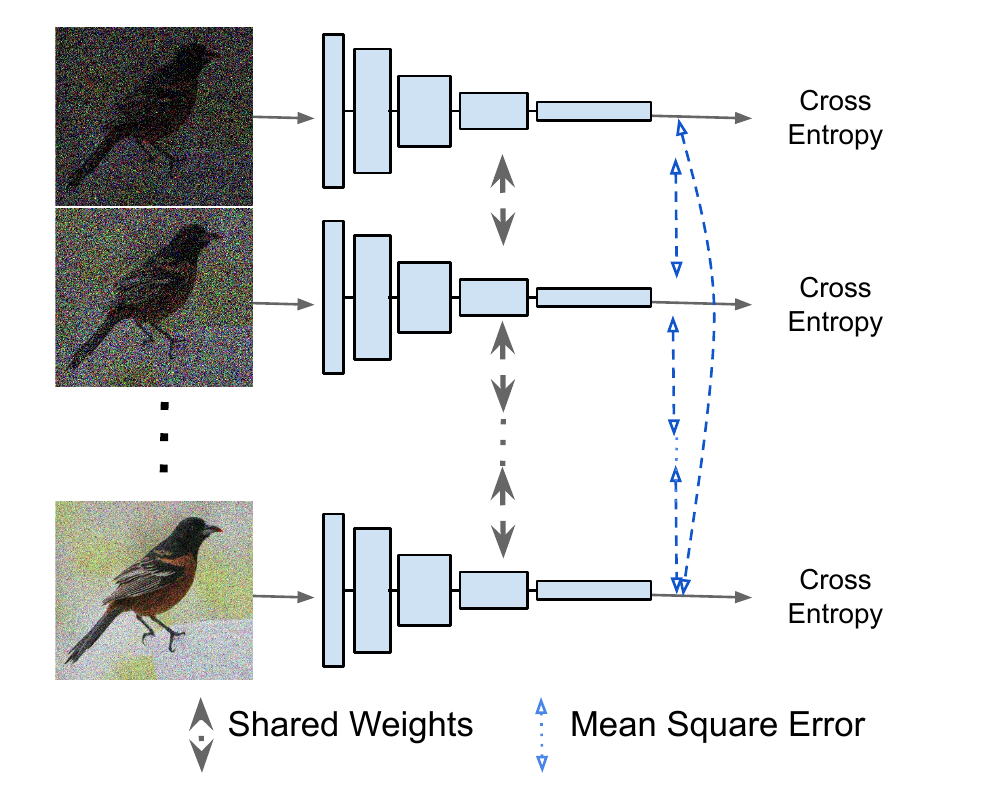}
    \vspace{-5pt}
    \caption{Photon Net (Ours)}
\end{subfigure}
\vspace{-5pt}
\caption{~\textbf{Comparison of Architectures with Existing Approaches for Low-Light Inference}: (a) \emph{Joint Denoising} consists of a denoiser jointly trained with an inference network. (b) \emph{Student Teacher Learning} uses a fixed teacher model trained on clean images and a trainable student model for noisy images. (c) \emph{Photon Net (Ours)} uses multiple images with different PPP level as input to the network. Different branches of network corresponds to different PPP levels, and all branches share weights with each other. A loss function such as mean squared error between feature representations is used to push images with different PPP level closer to each other in feature space.}
\label{fig:arch}
\vspace{-0.15in}\end{figure}

\subsection{Image Classification}
\noindent \textbf{Datasets:}
\label{sec:simulateddataset}
We first show the performance of our approach using simulated images using two datasets: CUB-200-2011 image classification dataset ~\cite{WahCUB_200_2011} and CARS dataset ~\cite{KrauseStarkDengFei-Fei_3DRR2013}. CUB-200-2011 is commonly used for fine-grained image classification benchmarks and consists of 200 species of birds with 5,994 training images and 5,794 test images. The CARS dataset contains images of 196 classes of cars (with different make, model and year) with 8,144 training images and 8,041 testing images.

We generate sequences of binary frames from the original images in the dataset (images captures by CMOS cameras) for training using the image formation model in Equation ~\ref{eq:imagingmodel}. $\phi \tau \eta$ in the model corresponds to the poisson parameter for the model. We assume $\eta*\tau = 1/1000$ to be constant for the dataset to simulate low flux setting and $\phi$ proportional to pixel value of original image. We then sum $N$ binary frames together to generate $\mathcal{S}^N$ images for the complete dataset. Figure ~\ref{fig:samplesimulatedimages} shows sample clean images from the dataset and sample noisy images generated using imaging model for Single Photon Cameras.

\smallskip
\noindent \textbf{Comparisons and Baselines:} \label{sec:comparisonapproaches}
We compare our method with two approaches that are designed for inference on low-SNR images. 
Our baseline approach is Joint Denoising~\cite{diamond2017dirty}, which uses a denoiser for noisy images coupled with a conventional image classification architecture. Both denoiser and classifier are trained together on noisy data (Fig.~\ref{fig:dirtypixelarch}), with a combined loss consisting of cross entropy loss for the classification and mean squared error for the denoiser.
We also compare our method with the Student-Teacher learning approach~\cite{gnanasambandam2020image} where clean images are used to train a \emph{fixed} teacher network and noisy images are used for training the student network (Fig.~\ref{fig:perceptarch}).
This approach encourages consistency between feature representation of clean image and noisy images by minimizing a mean squared error between the feature outputs of the student and the teacher networks. For more comparisons, please refer to the technical report.

\smallskip
\noindent\textbf{Experiments:}
\label{sec:implementation}
We perform all of our experiments with ResNet-18 ~\cite{he2016deep} as the backbone architecture provided by Pytorch ~\cite{NEURIPS2019_9015} for all the baselines. 
As shown in Figure ~\ref{fig:arch}, output of global pooling layer of size 512 is used as our feature extractor. 
We choose $5$ levels of Photon Scale Space images for training Photon Net in our experiments. This choice of number of levels is analysed later in the paper as part of an ablation study. All photon scale space images corresponding to the same image are sampled together in the same minibatch during training. We initialize our network with pre-trained weights from the model trained on clean images. Stochastic gradient descent with momentum optimizer with momentum as 0.9, base learning rate of 0.1 with cosine decay and batch size of 80 is used for fine-tuning.

\smallskip
\noindent\textbf{Results and Implications:} Table ~\ref{tab:clsresults} shows the results of our approach on CUB-200-2011 and CARS-196 dataset for different illumination levels. The proposed approach significantly outperforms Joint Denoising since denoising in the image space is not very effective for extreme noise levels (e.g., PPP $< 0.1$). With as few as $\sim 0.1$ Photons Per Pixel, our approach is able to get top-1 accuracy comparable to what conventional denoising approaches can achieve with 1 Photon Per Pixel (1 magnitude higher).
Student Teacher Learning performs better than Joint Denoising as it enforces feature consistency between noisy and clean images. However, since it uses a fixed teacher network with only clean images, the guidance is not as effective. Photon Net trains a wide gamut of SNR images together in the same network with noisy images.

\begin{table*}
\begin{center}
\begin{tabular}{c c c c c c c c c}
\hline
&& \multicolumn{3}{c}{{CUB-200-2011}} && \multicolumn{3}{c}{{CARS-196}}\\
\cline{3-5}\cline{7-9}
Test & PPP & Joint  & Student-Teacher & Photon Net && Joint & Student-Teacher & Photon Net \\
Data & & Denoising & Learning & (Ours) && Denoising & Learning & (Ours) \\
\hline
$\mathcal{S}^1$ & 0.11 & 27.21 & 35.43 & \textbf{42.37} && 34.51 & 57.81 & \textbf{64.23}\\
$\mathcal{S}^2$ & 0.22 & 31.33 & 39.50 & \textbf{48.56} && 43.14 & 65.85 & \textbf{70.51}\\
$\mathcal{S}^5$ & 0.53 & 39.41 & 44.46 & \textbf{55.19} && 57.11 & 71.13 & \textbf{75.23}\\
$\mathcal{S}^{10}$ & 1.07 & 44.17 & 48.08 & \textbf{58.68} && 65.78 & 73.51 & \textbf{78.97}\\
\hline
\end{tabular}
\end{center}
\vspace{-10pt}
\caption{\textbf{Image Classification Results}: Top-1 Accuracy results on CUB-200-2011 and CARS-196 dataset. Photon Net outperforms both Joint Denoising ~\cite{diamond2017dirty} and Student-Teacher Learning ~\cite{gnanasambandam2020image} on all noise levels.}
\label{tab:clsresults}\vspace{0pt}
\end{table*}

\smallskip
\noindent \textbf{Ablation Study:} We study the effect of the parameters of the Photon Scale Space (PSS) on the performance by varying the number of levels of PSS during training. We start with 2 levels of PSS (only noisy and clean image) and increase up to 9 (noisy, clean and 7 more intermediate levels). Fig. ~\ref{fig:ablationfactors} shows results of image classification on $\mathcal{S}^{1}$ test images of CUB-200-2011 and CARS-196 dataset. For these datasets, the performance of the model increases with increasing number of PSS levels, but saturates around 5 levels, thus informing the choice of parameters in our experiments. An important next step is to perform a similar empirical analysis for a wider range of datasets and tasks.

%% file: 06-monodepthestimation.tex
\subsection{Monocular Depth Estimation}

\begin{table*}
\begin{center}
\begin{tabular}{c c c c c c c c c}
\hline
Test DataSet & PPP & Method & $\delta_1\uparrow$  & $\delta_2\uparrow$ & $\delta_3\uparrow$ & rel$\downarrow$ & rms$\downarrow$ & $log_{10}\downarrow$\\
\hline
$\mathcal{S}^1$&0.11& Joint Denoising & 0.671 & 0.896 & 0.967 & 0.209 & 1.412 & 0.087\\
&& Photon Net (Ours) &\textbf{0.713} & \textbf{0.917} & \textbf{0.976} & \textbf{0.183} & \textbf{1.275} & \textbf{0.078}\\
\hline
$\mathcal{S}^{10}$ & 1.07 &  Joint Denoising & 0.763 & 0.941 & 0.984 & 0.162 & 1.177 & 0.069\\
 & & Photon Net (Ours) & \textbf{0.793} & \textbf{0.953} & \textbf{0.987} & \textbf{0.149} & \textbf{1.104} & \textbf{0.063}\\
\hline
\end{tabular}
\end{center}
\vspace{-15pt}
\caption{\textbf{Monocular Depth Estimation Results:} on NYUV2 dataset.}
\label{tab:monodepthresults}
\vspace{-0.1in}
\end{table*}




We also show the application of our approach to monocular depth estimation, a regression task. \smallskip

\noindent \textbf{Depth Estimation Overview:} For this application, we use the DenseDepth~\cite{alhashim2018high} base architecture, consisting of an Encoder and a Decoder. The Encoder is a deep CNN (ResNet-34 pretrained on ImageNet~\cite{deng2009imagenet}) which extracts the feature maps and the Decoder is a series of upsampling layers with skip connections to construct the depth map from the feature maps. Loss function used is a combination of point-wise L1 loss and Structural Similarity loss between predicted and ground truth depth values. We use the same training procedure as described in~\cite{alhashim2018high}. \smallskip

\noindent \textbf{Photon Net training for Depth Estimation:}
We train our Depth Estimation architecture with photon scale space images. Mean Squared Error Loss is used for feature consistency of the feature outputs of the images from different PPP levels. We use output of the Encoder network (after global pooling layer) for our feature representation. We provide more details on the architecture in the technical report.

\smallskip
\noindent \textbf{Experiments and Results}
We evaluate our approach on NYUV2 dataset ~\cite{silberman2012indoor}. Same training and testing split is used as ~\cite{alhashim2018high} which includes 50K training and 654 testing images. We simulate SPC images using the same procedure described earlier in Section ~\ref{sec:simulateddataset}.
The following six standard evaluation metrics are used:
\vspace{-5pt}
\begin{itemize}
\itemsep -2pt 
\item average relative error (rel):  $\frac{1}{n} \Sigma^n_p \frac{y_p - \hat{y}_p}{y}$, 
\item root mean square error (rms): $\sqrt{ \frac{1}{n} \Sigma^n_p (y_p - \hat{y}_p)^2} $, 
\item average ($log_{10}$) error: $\frac{1}{n} \Sigma^n_p |log_{10}(y_p) - log_{10}(\hat{y}_p)|$ and
\item threshold accuracy ($\delta_i$): \% of $y_p$ s.t. max($\frac{y_p}{\hat{y}_p}$, $\frac{\hat{y_p}}{y_p}$) = $\delta_i < thr$ for $thr = 1.25$, $1.25^2$, $1.25^3$ 
\end{itemize}
\vspace{-5pt}
where $y_p$ is a pixel in depth image $y$, $\hat{y}_p$ is a pixel in the predicted depth image $\hat{y}$, and n is the total number of pixels for each depth image.

\begin{figure}
\begin{center}
\includegraphics[width=\linewidth]{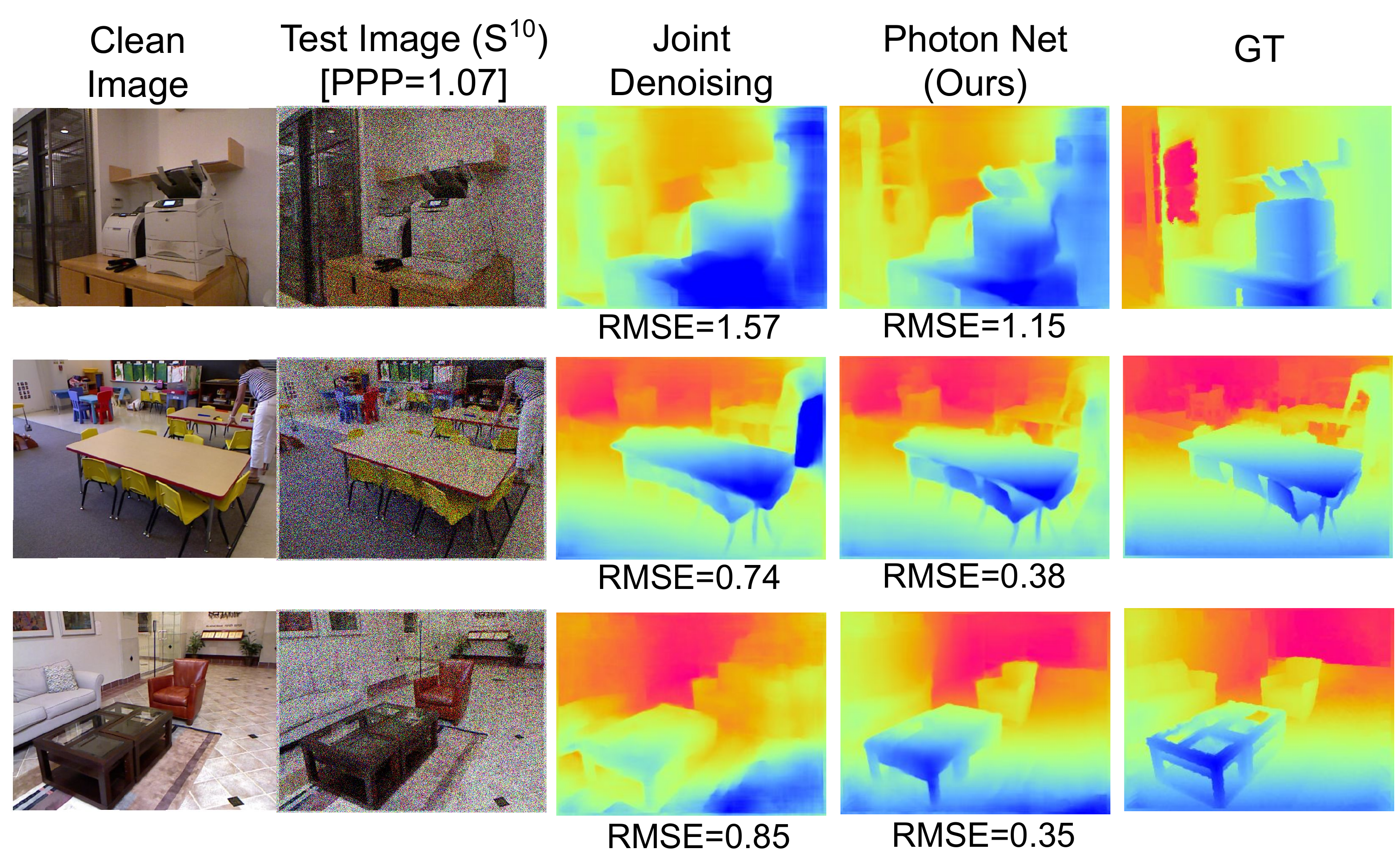}
\end{center}
\vspace{-15pt}
\caption{\textbf{Estimated depth maps}: Comparison of depth maps from Photon Net (Ours) and the baseline on NYUV2 test images $\mathcal{S}^{10}$ (PPP$\sim$1.07).}
\label{fig:visualmonodepth}
\vspace{-0.1in}\end{figure}

We compare our method with Joint Denoising, which uses a denoiser with Depth Estimation architecture.
Table~\ref{tab:monodepthresults} shows results on NYUV2 dataset and Fig.~\ref{fig:visualmonodepth} shows example depth output results of our approach and the baseline.
Photon Net outperforms baseline approach both qualitatively and quantitatively for multiple noise levels.

\begin{figure}
\begin{subfigure}{0.49\linewidth}
    \centering
    \includegraphics[width=\linewidth]{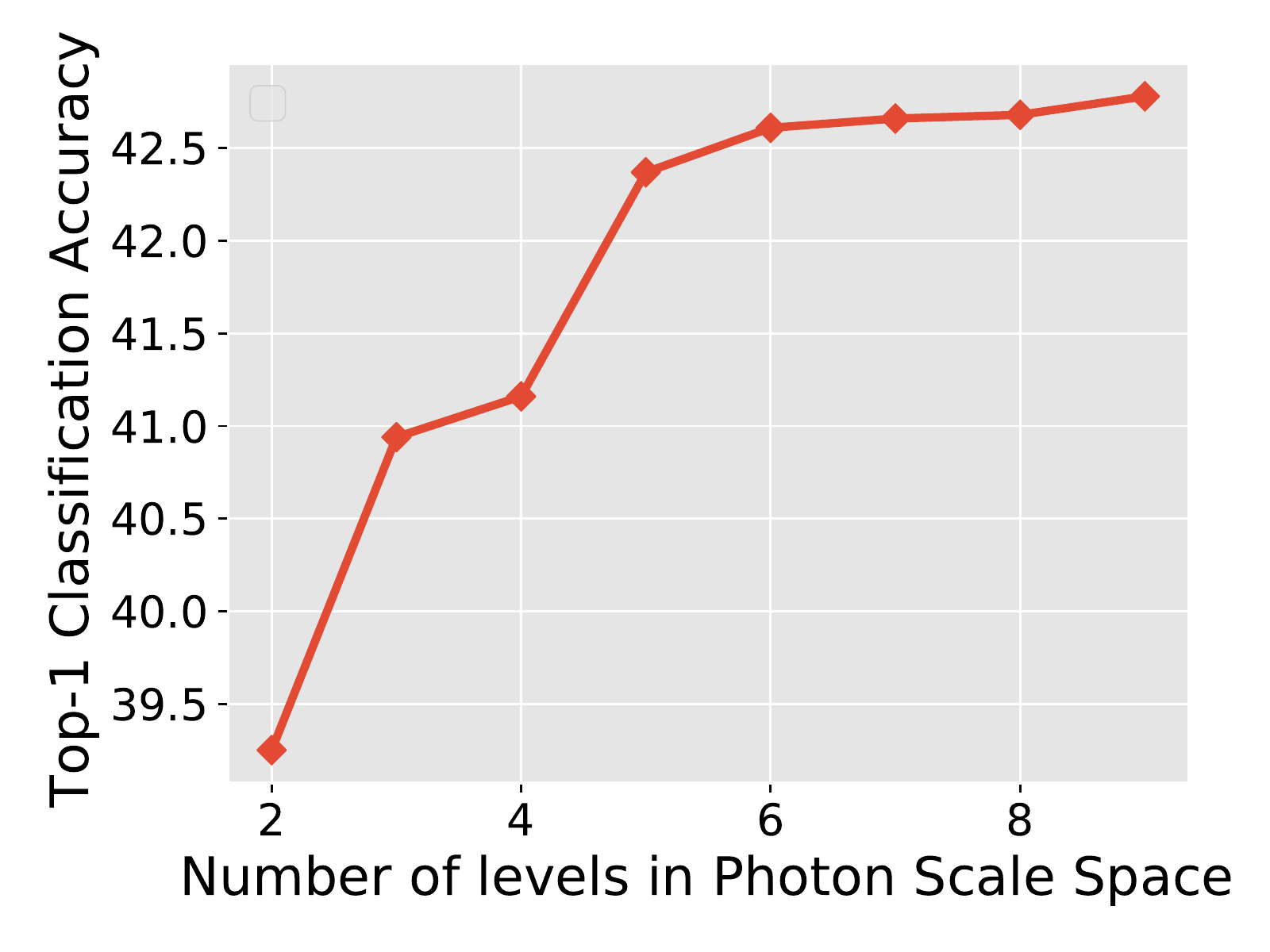}
    \caption{CUB-200-2011}
\end{subfigure}
\begin{subfigure}{0.49\linewidth}
    \centering
    \includegraphics[width=\linewidth]{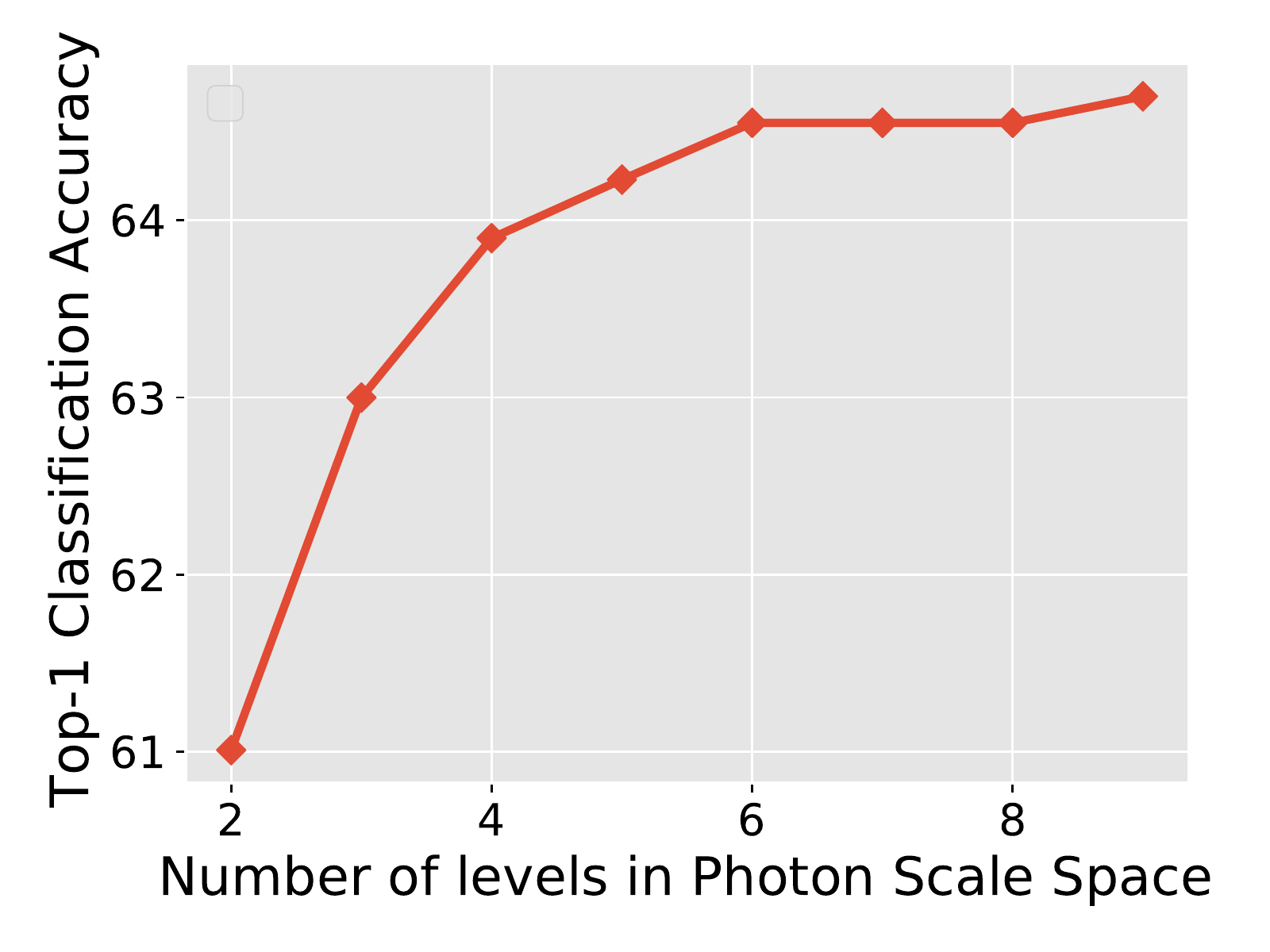}
    \caption{CARS-196}
\end{subfigure}
\vspace{-5pt}
\caption{{\bf Effect of Photon Scale Space Parameters on Inference Performance:} Top-1 classification accuracy of Photon Net on $\mathcal{S}^1$ test images (PPP$\sim$0.11) with increasing number of levels in Photon Scale Space. Performance increases with increasing number of levels in PSS and saturates at 5-6 levels for both datasets.}
\label{fig:ablationfactors}
\end{figure}

%% file: 07-realexperiments.tex
\section{Experiments on Real SPAD Images}
\label{sec:realexperiments}
\begin{figure}
\centering
\begin{subfigure}{0.4\linewidth}
    \includegraphics[width=\linewidth]{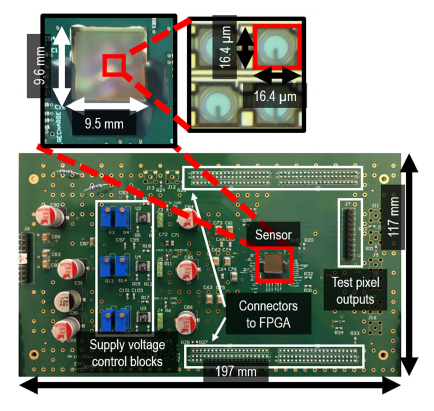}
\end{subfigure}
\begin{subfigure}{0.5\linewidth}
    \includegraphics[width=\linewidth]{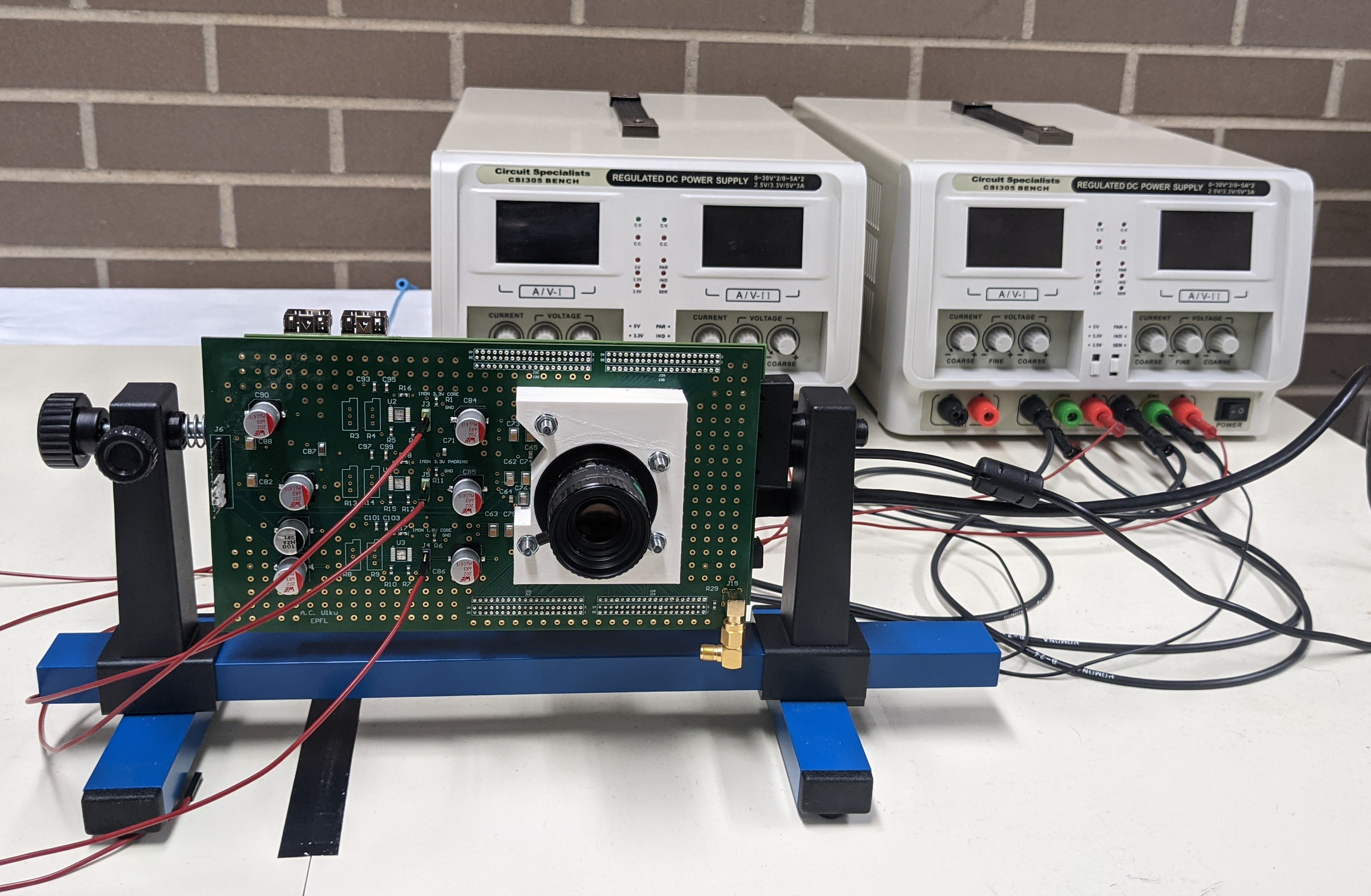}
\end{subfigure}
\vspace{-5pt}
\caption{\textbf{Camera Setup:} SwissSPAD2 board (Left) and the setup for image capture (Right).}
\label{fig:setup}
\end{figure}

In order to evaluate the validity of the SPC image simulation model and the proposed approaches on real SPAD images, we collect a data-set of SPAD images using a SwissSPAD2 camera~\cite{ulku2018512} (Fig.~\ref{fig:setup}). 

\smallskip
\noindent \textbf{Camera Setup:} We operate the camera in the binary mode where it captures binary frames at a spatial resolution of 512×256 with maximum frame rate at 96.8kHz. Currently, the sensor is not equipped with Bayer filters, so only gray-scale (single channel) frames are captured. The captured images contain hot pixels which we correct in post processing. We capture an image of a black scene to identify the location of the hot pixels and then filter them by using spatial neighborhood information. 

\begin{figure}
    \centering
    \includegraphics[width=\linewidth]{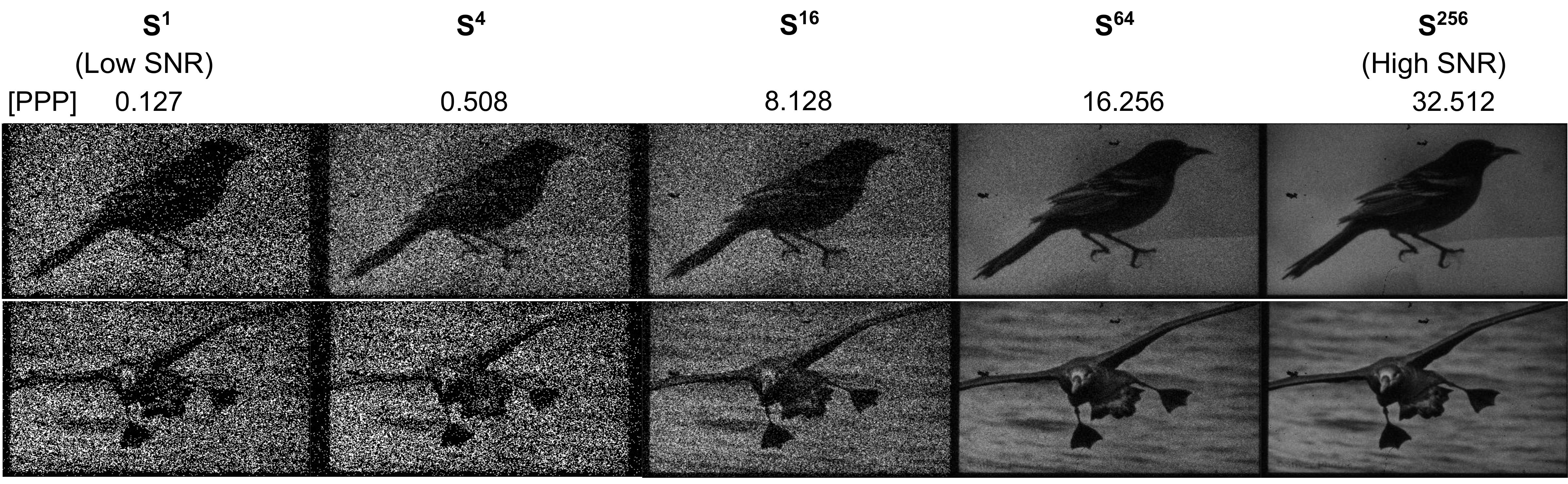}
\caption{\textbf{Real SPAD images} captured from CUB-200-2011 dataset using the SwissSPAD2 camera.}
\label{fig:samplespadimages}
\end{figure}

\begin{figure}
    \centering
    \includegraphics[width=\linewidth]{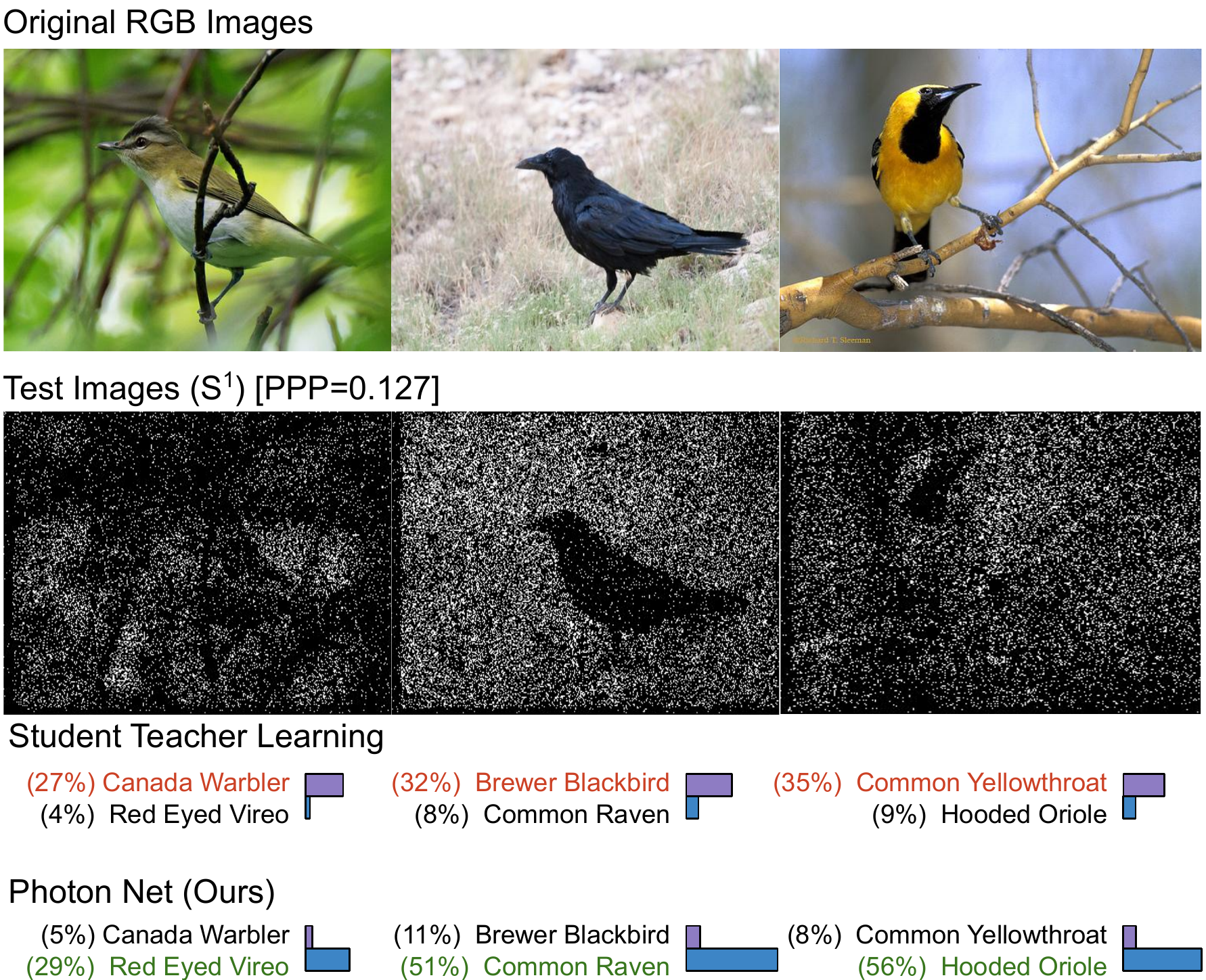}
\caption{\textbf{Results with Real SPAD Sensor} of image classification on CUB-200 dataset for $\mathcal{S}^1$ test images with prediction probabilities output by both Student Teacher Learning and Photon Net (Ours). Classification output is highlighted in red for wrong prediction and green for correct prediction.}
\label{fig:spadresults}
\end{figure}

\smallskip \noindent \textbf{Dataset:} For the dataset collection for image classification task, we displayed the original RGB images on a monitor screen (Dell P2419H, 60Hz) and then captured it using SPAD sensors. The camera is placed at around 1m distance from the screen and positioned to cover the display in its field of view. We selected a subset of images from CUB-200-2011 dataset (CUB-subset) for the data collection, including 3656 training images and 3518 testing images from a randomly collected subset of 122 categories. Fig. ~\ref{fig:samplespadimages} shows examples of $N$-Sum images captured by the camera.

\smallskip \noindent \textbf{Experiments and Results:}
We follow the same procedure for training as described in Section ~\ref{sec:implementation}.
Table ~\ref{tab:realclsresults} shows results of our approach on real images from SPADs. Although the overall accuracy levels are lower (for all approaches) than those with simulated images due to the real images having a lower resolution and only gray-scale intensities (no Bayer filter on the real SPAD sensor), Photon Net outperforms both baselines on all noise levels. 

Fig.~\ref{fig:spadresults} shows output probabilities of the predicted classes with ground truth for a few samples. Even in extreme low-light conditions with PPP as low as $~\sim 0.1$, the proposed photon net approach is able to recover correct class labels. 

\begin{table}
\begin{center}
\begin{tabular}{c c c c c}
\hline
\small{Test} & \small{PPP} & \small{Joint}  & \small{Student-Teacher} & \small{Photon Net} \\
\small{Data} & & \small{Denoising} & \small{Learning} & \small{(Ours)}\\
\hline
$\mathcal{S}^1$ & 0.127 & 13.34 & 17.54 & \textbf{21.78}\\
$\mathcal{S}^2$ & 0.254 & 16.57 & 20.67 & \textbf{26.74}\\
$\mathcal{S}^4$ & 0.508 & 18.82 & 24.55 & \textbf{32.33}\\
$\mathcal{S}^{8}$ & 1.016 & 21.07 & 28.34 & \textbf{35.79}\\
$\mathcal{S}^{16}$ & 2.032 & 24.91 & 29.82 & \textbf{39.14}\\
\hline
\end{tabular}
\end{center}
\vspace{-10pt}
\caption{\textbf{Experiments with real SPAD data.} Top-1 image classification results on CUB-subset images captured using a SPAD camera.}
\label{tab:realclsresults}
\vspace{-0.2in}\end{table}

%% file: 08-conclusion.tex
\section{Discussion and Limitations}
\noindent {\bf Low-light inference beyond classification and depth estimation:} So far, we have demonstrated the benefits of the proposed approaches for image classification and depth estimation tasks. A natural direction is to extend these ideas to inference models for a larger gamut of image inference and scene understanding tasks, including object detection~\cite{ren2015faster}, instance segmentation~\cite{he2017mask} and key-point detection~\cite{newell2016stacked}\smallskip

\noindent {\bf Inference in high-flux scenarios:} Although the primary focus of this paper is on low-light inference, due to the high dynamic range capabilities of SPADs~\cite{antolovic_dynamic_2018,Ingle:2019,ma2020quanta}, the proposed techniques can be adapted for inference in extremely bright scenes where conventional sensors get saturated. \smallskip


\noindent{\bf Inference on time-varying inputs:} In their current form, the proposed approaches assumes static single-frame input.  However, most current single-photon sensors~\cite{ma2017photon,ulku2018512} can capture binary frames at high speeds, up to several thousand frames per second. A promising future research direction is to perform inference in the presence of high-speed camera / scene motion on a temporal sequences of such low bit-depth frames, while exploiting temporal correlations.

\noindent{\bf Acknowledgement.} This research was supported in parts by the NSF CAREER Award 1943149, a Sony Faculty Innovation Award, NSF Award 2003129, and the Intel MLWiNS Award. We also thank Edoardo Charbon for providing access to the SwissSAPAD2 array used for the experiments and Sizhuo Ma for helping with the image capturing setup and data collection using the array.


%% file: 09-techreport.tex
In this report, we provide technical details and results that are not included in the main paper due to space constraints.

\subsection{Image Classification}

\subsubsection{Architecture Overview}

\begin{figure}[H]
\begin{subfigure}{\linewidth}
    \centering
    \includegraphics[width=0.6\linewidth]{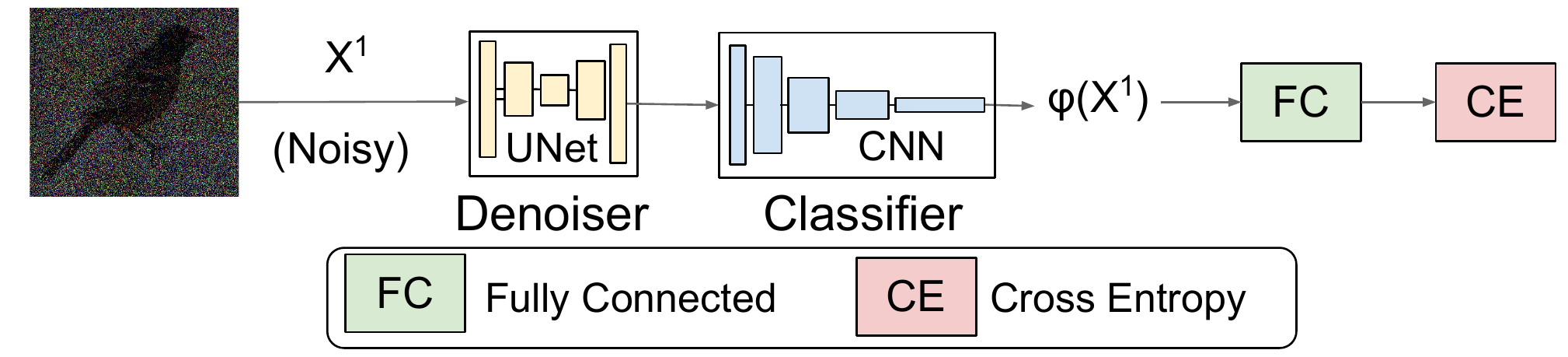}
    \caption{Joint Denoising}
    \label{fig:detailedjoint}
\end{subfigure}
\begin{subfigure}{\linewidth}
    \centering
    \includegraphics[width=0.6\linewidth]{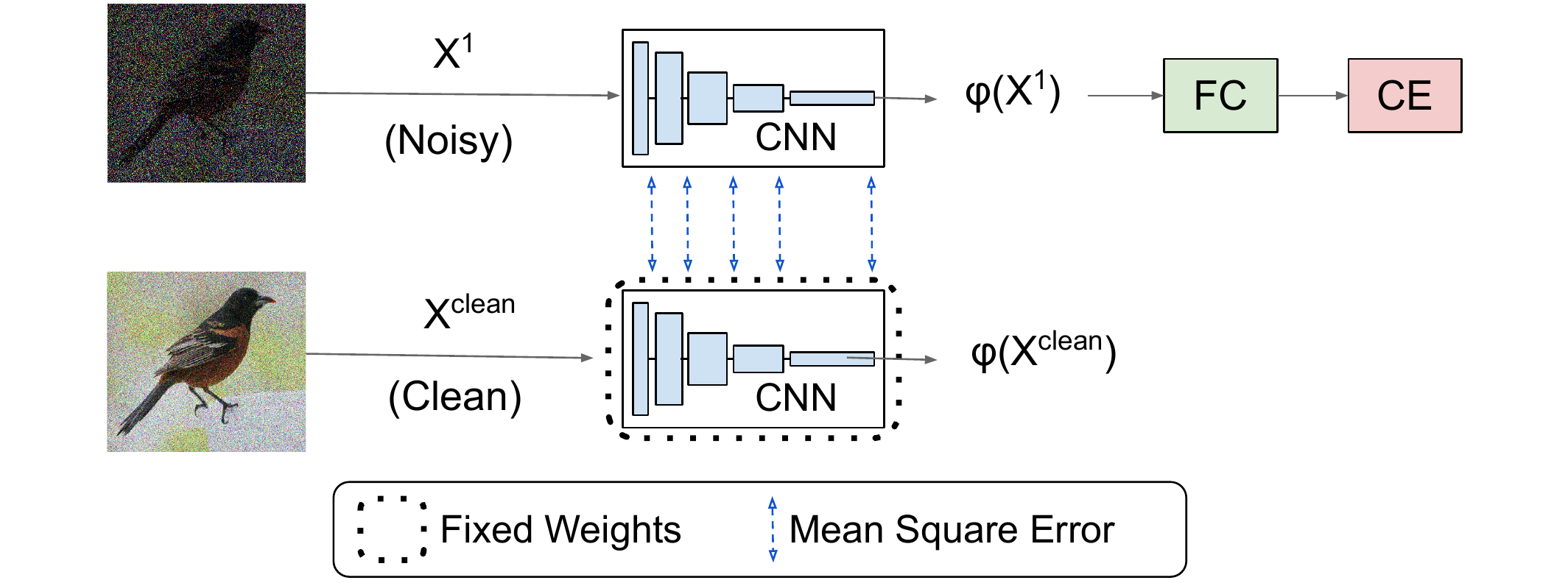}
    \caption{Student Teacher Learning}
    \label{fig:detailedstudent}
\end{subfigure}
\begin{subfigure}{\linewidth}
    \centering
    \includegraphics[width=0.6\linewidth]{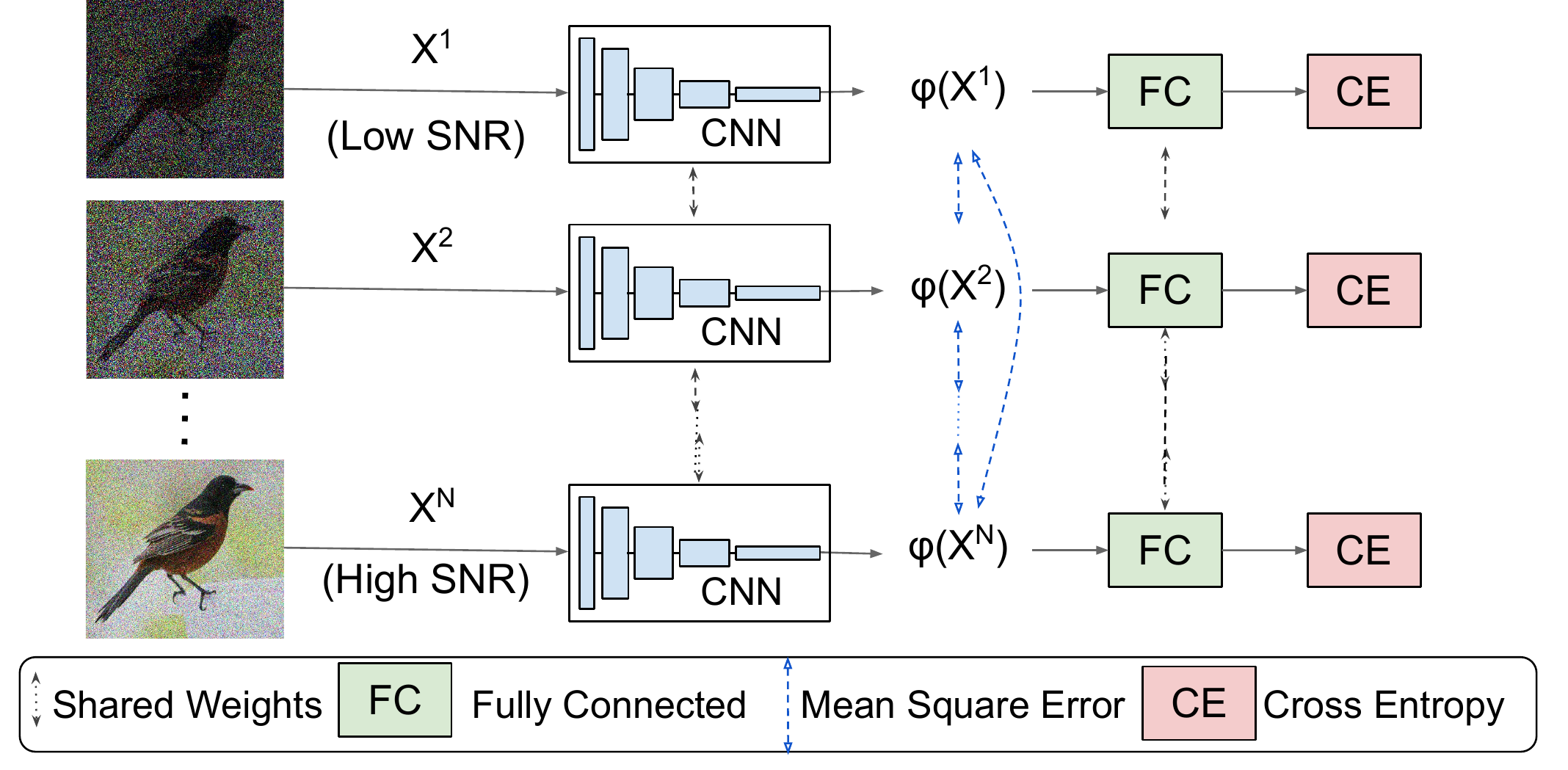}
    \caption{Photon Net (Ours)}
\end{subfigure}
\caption{\textbf{Architecture Overview} for Image Classification}
\label{fig:detailedarch}
\end{figure}

\noindent We provide more detailed overview of the architectures used for the approaches used for image classification task.\smallskip

\noindent \textbf{Joint Denoising}
Joint Denoising architecture ~\cite{diamond2017dirty} consists of a joint network with a denoiser (20 layer UNet) and a CNN classifier (Resnet-18 ~\cite{he2016deep}). We use Mean Squared Error loss for the denoiser which uses noisy and clean images. Cross Entroy Loss is used for the classifer with uses the class label of the image. The joint network is trained with sum of both the losses (Figure ~\ref{fig:detailedjoint}).
The denoiser is initialized with pretrained weights on noisy and clean images.

\noindent \textbf{Student Teacher Learning}
Student Teacher architecture ~\cite{gnanasambandam2020image} is composed of a teacher network and a student network. Teacher network (ResNet-18) is a pre-trained classifier on clean images. Student Network uses the same network architecture as the teacher network (ResNet-18). Intermediate feature output maps ('relu', 'layer1', 'layer2', 'layer3', 'layer4' from pytorch's implementation) from the CNN Network of both student and teacher network is used for feature consistency. Final training consists of training the student network with cross entropy loss and mean squared error loss while teacher network is kept fixed (Figure \ref{fig:detailedstudent}).
Student Teacher learning uses double the network parameters for classifier during training but only uses student network for testing.

\noindent \textbf{Photon Net (Ours)}
Photon Net training uses multiple images with different PPP level as input to the network. Different branches of the network are CNN architectures (ResNet-18) which share weights with each other and act as a feature extractor. Images with different PPP levels are sampled together in the same mini-batch so gradients from high SNR image branches can guide the low SNR images.
The feature output from the final layer (after global pooling layer) is used for the feature consistency of different PPP levels using Mean Square Error Loss.
Cross Entropy Loss is used for the training the image classifier which uses the classification label.

\subsubsection{Additional Results}

\begin{figure}[H]
\begin{center}
\includegraphics[width=0.5\linewidth]{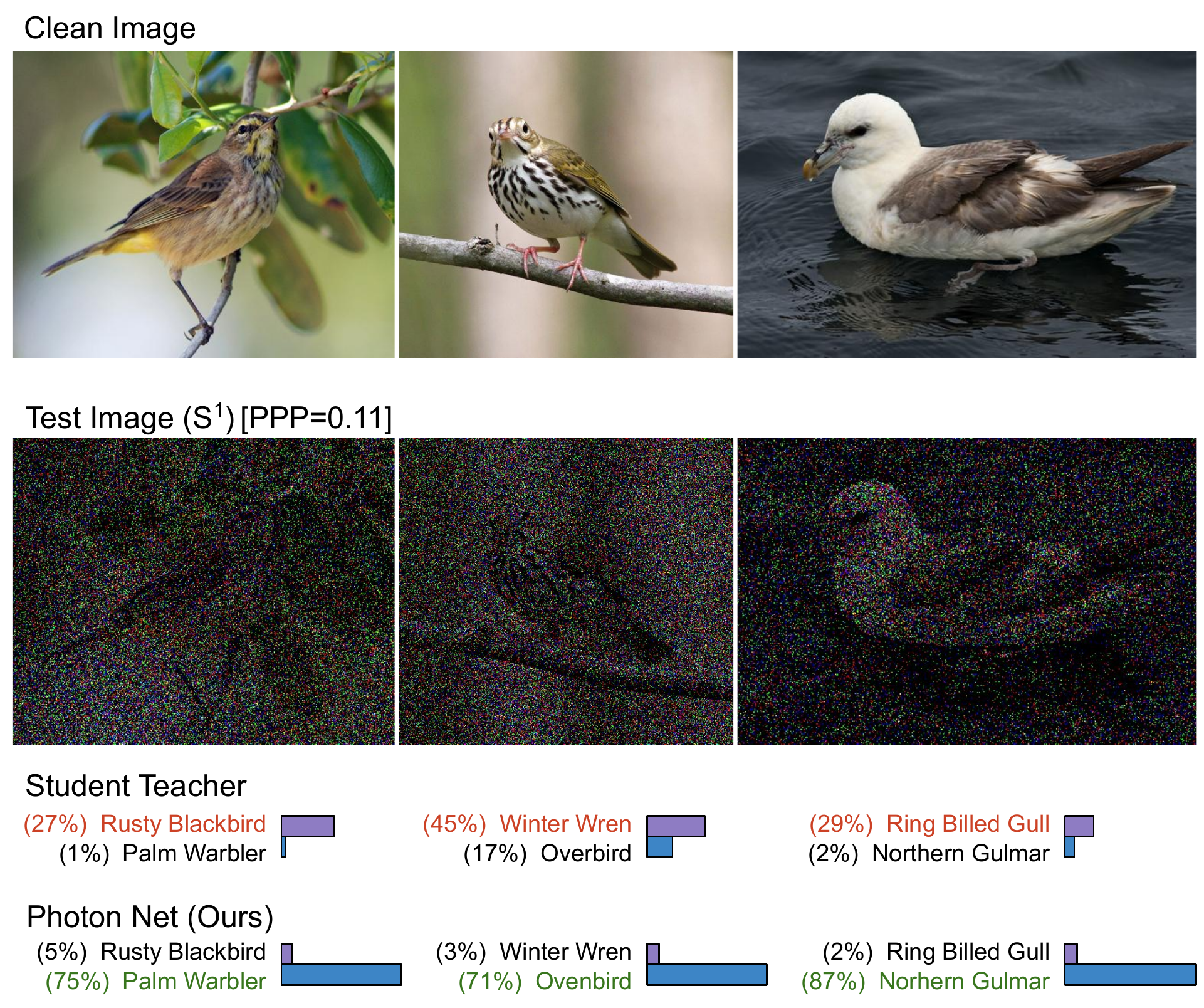}
\end{center}
\caption{\textbf{Image Classification Results} using Photon Net on CUB-200-2011 Dataset for $\mathcal{S}^1$ test images.}
\label{fig:suppvisualcompcub}
\end{figure}

\begin{figure}[H]
\begin{center}
\includegraphics[width=0.5\linewidth]{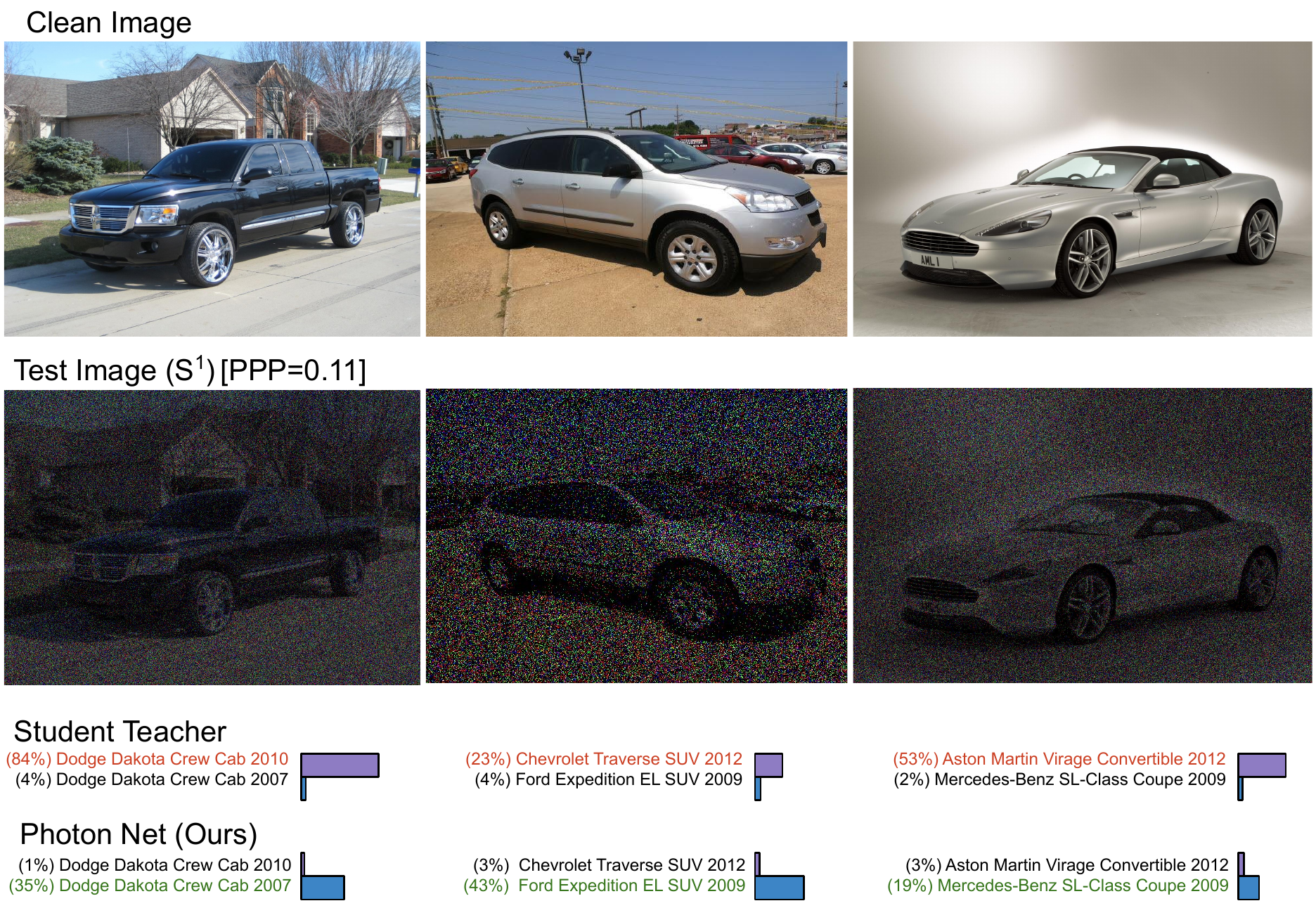}
\end{center}
\caption{\textbf{Image Classification Results} using Photon Net on CARS Dataset for $\mathcal{S}^1$ test images.}
\label{fig:suppvisualcompcars}
\end{figure}

\begin{figure}[H]
\begin{center}
\includegraphics[width=0.5\linewidth]{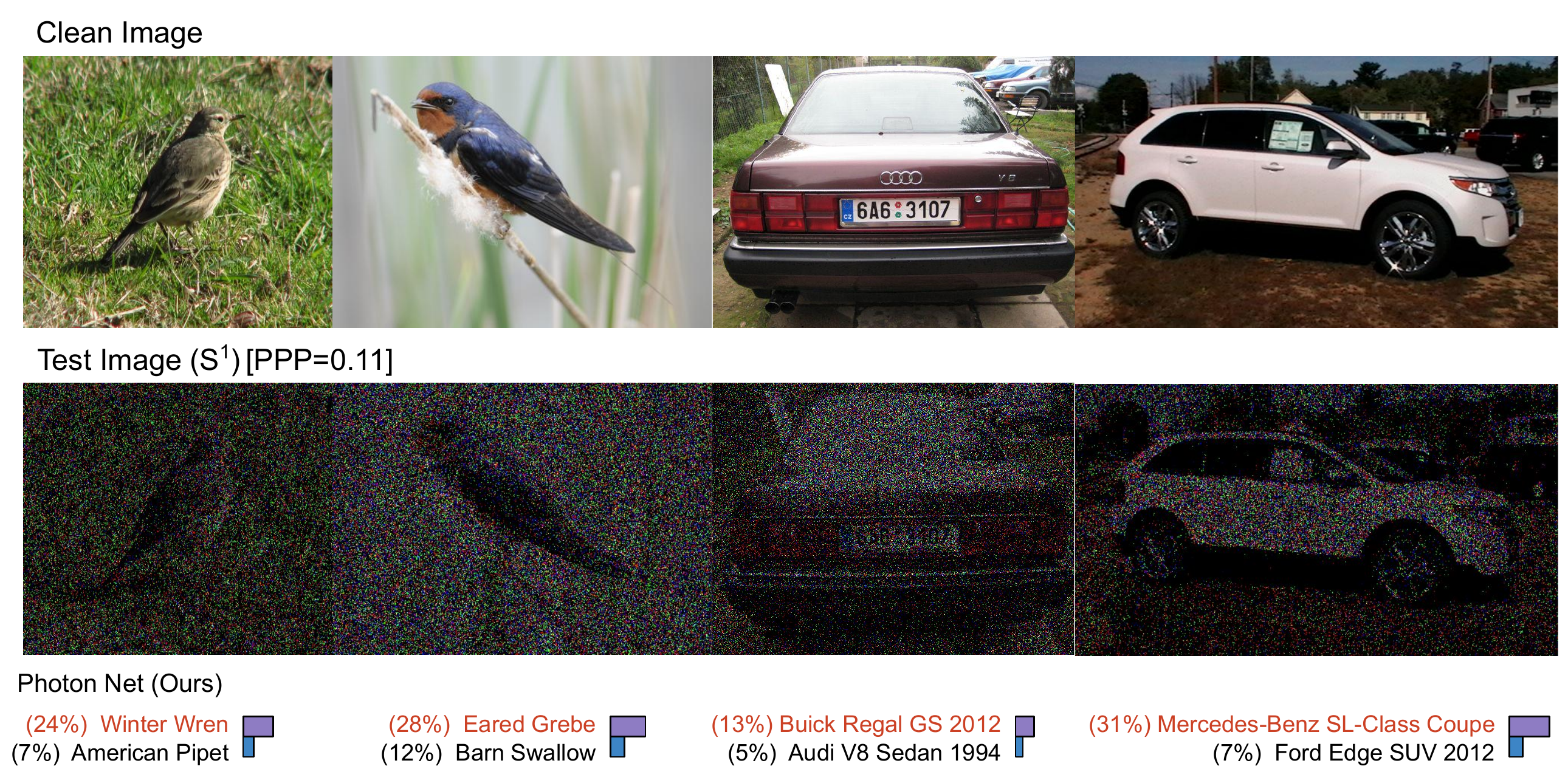}
\end{center}
\caption{\textbf{Few Failure cases examples} of Photon Net on CUB-200-2011 and CARS dataset for $\mathcal{S}^1$ test images.}
\label{fig:suppfailurecases}
\end{figure}

Figure ~\ref{fig:suppvisualcompcub} and ~\ref{fig:suppvisualcompcars} shows results of image classification on CUB-200-2011 ~\cite{WahCUB_200_2011} and CARS ~\cite{KrauseStarkDengFei-Fei_3DRR2013} dataset $\mathcal{S}^1$ test images using Photon Net. Probability output of incorrect class is highlighted in red and correct class is highlighted in green. Even in the case of extreme low light (PPP$~$0.1), Photon Net is able to recover the correct output label. Figure ~\ref{fig:suppfailurecases} example of few failure cases where Photon Net architecture fails to get the correct prediction. As we can observe, these cases are extremely challenging.

\subsubsection{More ablation studies}
\begin{figure}[H]
\centering
\begin{subfigure}{0.3\linewidth}
    \centering
    \includegraphics[width=\linewidth]{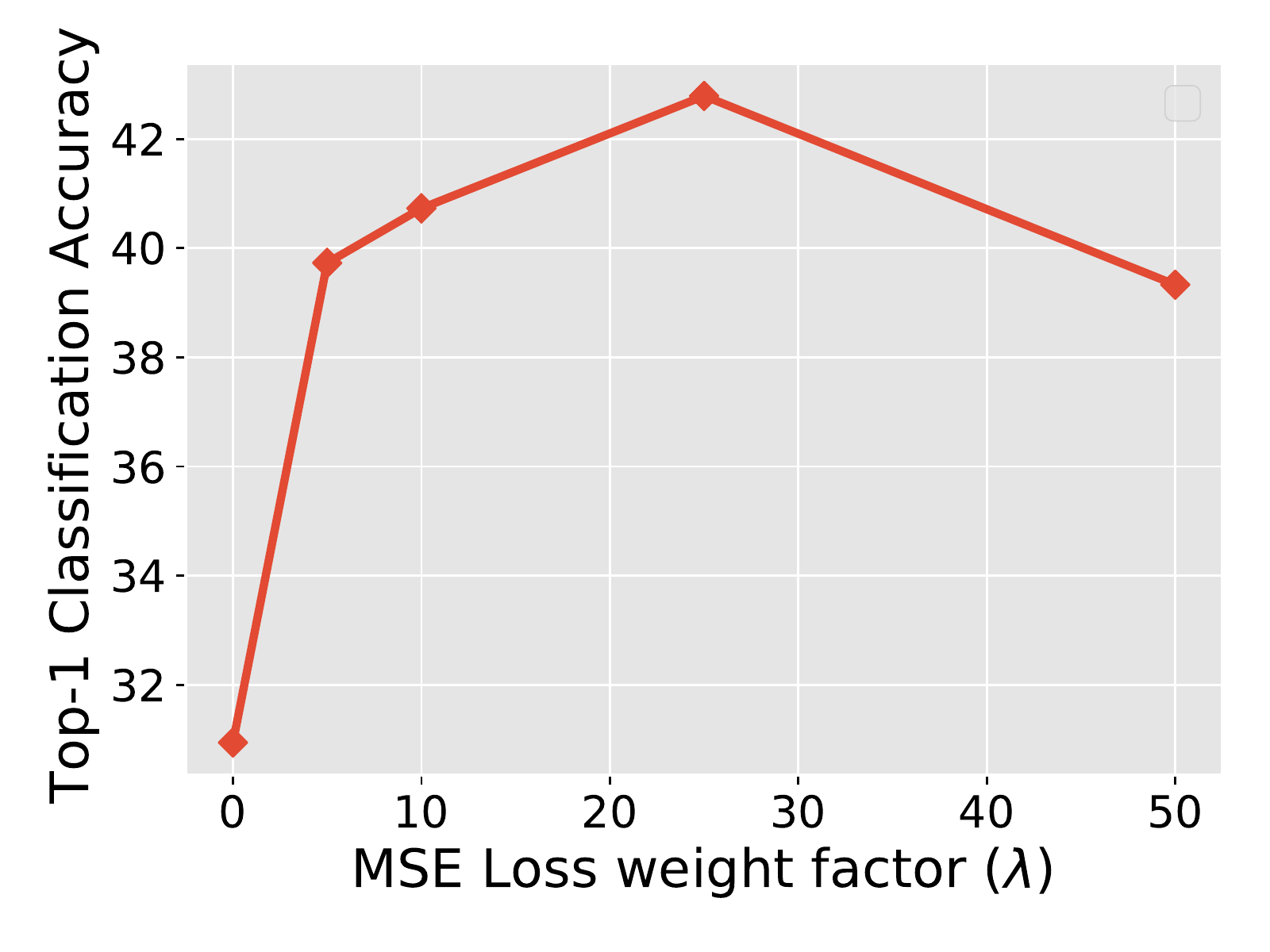}
    \caption{MSE Loss weight factor ($\lambda$)}
\end{subfigure}
\begin{subfigure}{0.3\linewidth}
    \centering
    \includegraphics[width=\linewidth]{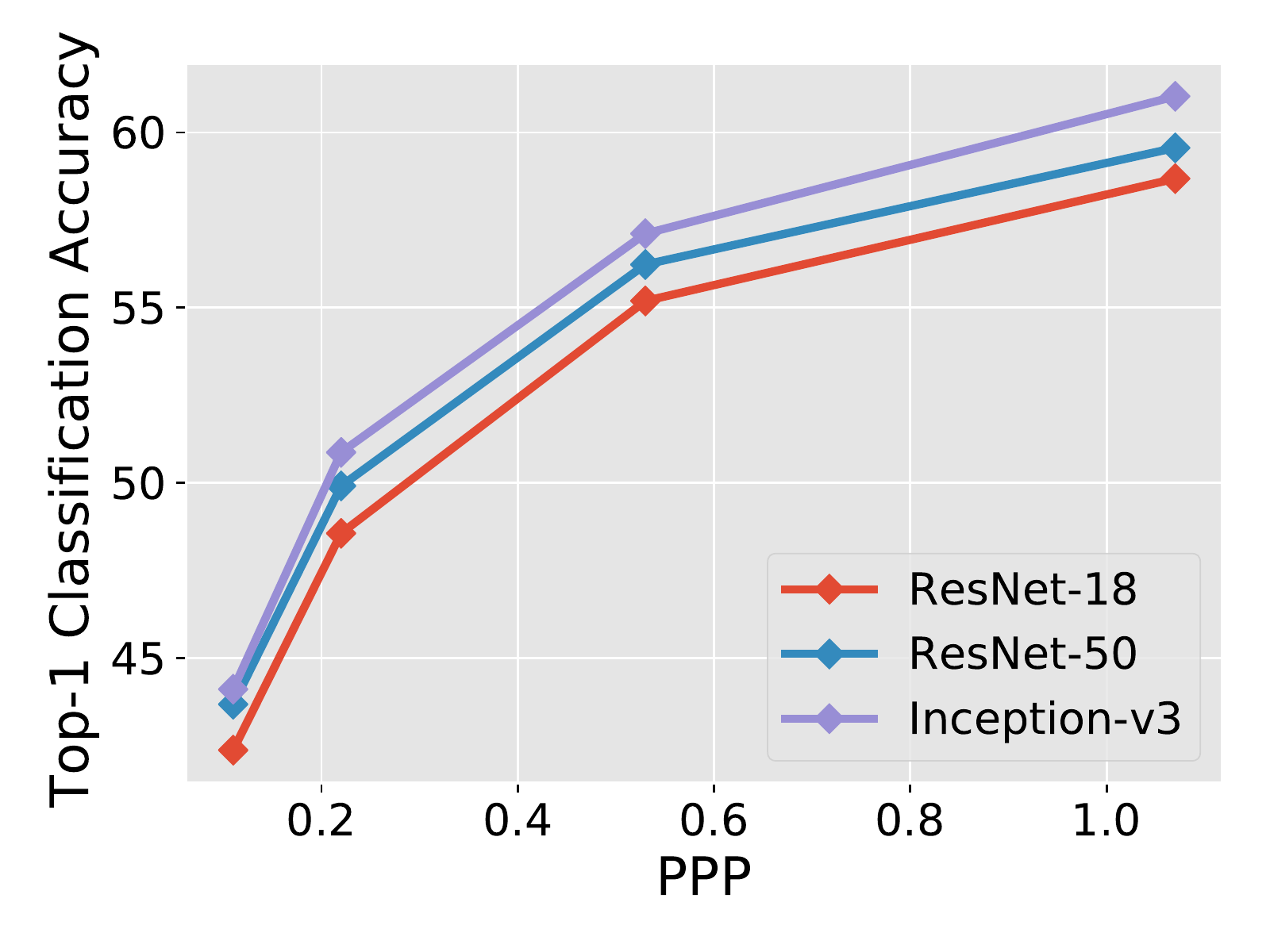}
    \caption{Different CNN architectures for Photon Net}
\end{subfigure}
\vspace{-5pt}
\caption{\textbf{Ablation Studies}: Performance of Photon Net training while varying: (a) MSE loss weight factor ($\lambda$) , (b) base architecture}
\label{fig:moreablation}
\vspace{-0.1in}\end{figure}

We study the effect of the hyper parameter of the Photon Net training on the performance. We vary the weighting factor of the MSE loss in the overall loss for image classification. We start with $\lambda$=0 and increase upto $\lambda=50.0$. Figure ~\ref{fig:moreablation} shows Photon Net performs best for $\lambda$=25.0.

We also analyse the performance of Photon Net using different base architecture for the feature extractor. We compare ResNet-18 with deeper CNN architectures such ResNet-50 and InceptionV3. ~\cite{szegedy2016rethinking}. Figure ~\ref{fig:moreablation} shows increase in the performance of Photon Net with deeper CNN architectures. This shows the versatility and ease to extend Photon Net to different CNN architectures.

\begin{table}[H]
\begin{center}
\begin{tabular}{c c c c c c c c}
\hline
Test & PPP & Vanilla Net & Vanilla Net w/ & BM3D & Curriculum & Student Teacher & Photon Net \\
Data && & Photon Scaled Images & Denoising & Learning & Learning (N-steps) &(Ours) \\
\hline
$\mathcal{S}^1$ & 0.11 & 21.35 & 28.92 &25.52& 33.72 & 35.79  & \textbf{42.37}\\
$\mathcal{S}^2$ & 0.22 & 25.61 & 34.51 &29.15& 39.44 & 42.16 & \textbf{48.56}\\
$\mathcal{S}^5$ & 0.53 & 37.14 & 43.26 &38.81& 44.99 & 46.91 & \textbf{55.19}\\
$\mathcal{S}^{10}$ & 1.07 & 42.99 & 44.63 &43.34& 48.65 & 48.86 & \textbf{58.68}\\
\hline
\end{tabular}
\end{center}
\vspace{-10pt}
\caption{\textbf{Ablation Study}: Top-1 Accuracy results of image classification on CUB-200-2011 dataset }
\label{tab:curriculumresults}\vspace{0pt}
\end{table}



We perform an ablation study to analyse the individual contribution of Photon Net training and using Photon Scaled Images in the final performance. Table ~\ref{tab:curriculumresults} shows Top-1 accuracy on CUB-200-2011 dataset. `Vanilla Net` represents the training procedure where a conventional image classification CNN model (ResNet-18) is trained with cross entropy loss using only noisy images. `Vanilla Net w/ Photon Scaled Images` trains the Vanilla Net with photon scaled images. As we an see, adding Photon Scale Space images increases the performance by about 8-9\% on all noise levels and shows the effectiveness of high SNR images in training. Photon Net training further improves the model by more than 13\% as feature consistency loss increases the robustness to noise. `BM3D denoising` shows the performance of Vanilla Net training on denoising training and testing images using BM3D algorirhtm.

We also compare our model to Curriculum Learning technique, where the Vanilla Net is trained in N steps, starting with only the clean images first step and successively finetuning the model by adding images with higher noise levels in next steps. Photon Net outperforms Curriculum Learning as it uses the high SNR as a guide more effectively by adding the feature consistency loss. We also do Student Teacher Learning in N-steps (N is number of photon scaled levels) using the Photon Scaled Images. We use successive levels of photon scaled images for student and teacher network. Photon Net performs better Student Teacher Learning by significant margin.

\subsection{Monocular Depth Estimation}
\subsubsection{Architecture Overview}

\begin{figure}[H]
\begin{subfigure}{\linewidth}
    \centering
    \includegraphics[width=0.6\linewidth]{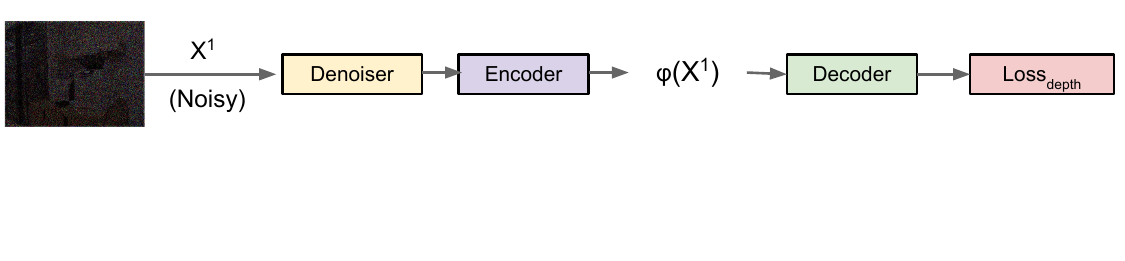}
    \vspace{-30pt}
    \caption{Joint Denoising}
\end{subfigure}
\begin{subfigure}{\linewidth}
    \centering
    \includegraphics[width=0.6\linewidth]{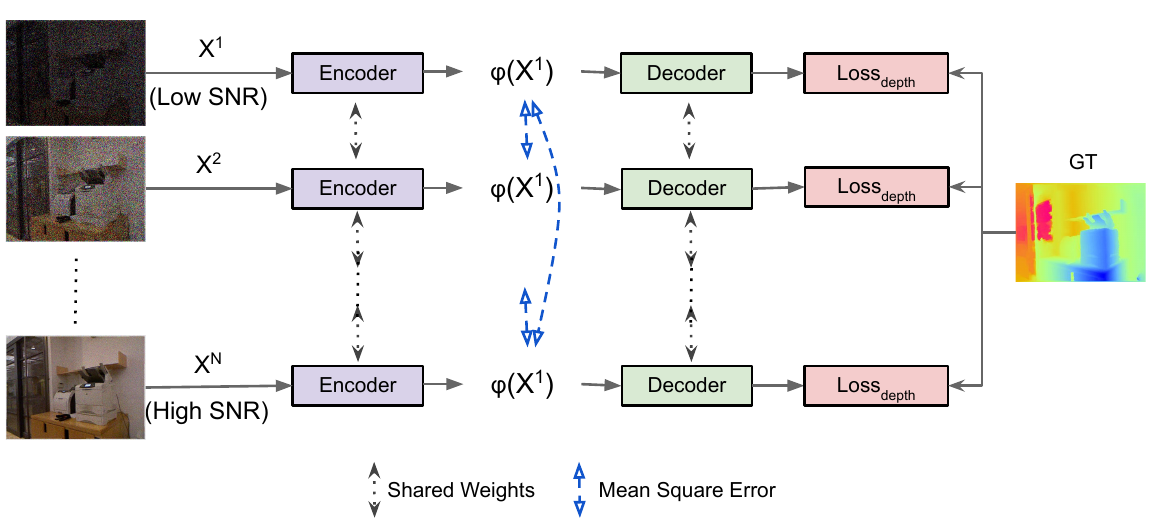}
    \caption{Photon Net (Ours)}
\end{subfigure}
\caption{~\textbf{Overview of the Depth Estimation with Photon Net}: }
\label{fig:detailedMDEarch}
\end{figure}
\noindent \textbf{Joint Denoising}
Joint Denoising consists of a depth estimation architecture based on DenseDepth ~\cite{alhashim2018high} coupled with a denoiser for noisy images. Denoiser is a UNet network (20 layers) which is pretrained on noisy and clean images using Mean Square Error Loss. DenseDepth architecture for depth estimation consists of an encoder network (Deep CNN network pretrained on Imagenet) and a decoder network (upsampling layers with skip connects) that generates the output depth maps. Loss function for depth estimation is a combination of point wise L1 loss and Structural Similarity loss between predicted and ground truth depth values. Overall Loss is the sum of losses from denoiser and depth estimation.\smallskip

\noindent \textbf{Photon Net}
Photon Net architecture takes multiple images with different PPP levels as the input to the network. Different branches of the network are the encoder networks with shared weights. We use the same encoder and decoder as baseline for fair comparison. Different images are sampled together in the same mini-batch in order for high SNR images to guide the low SNR images. Final feature output map from the encoder (after global pooling layer) is used for the feature consistency of different PPP levels (using Mean Square Error Loss). Overall Loss is the combination of Mean Square Error loss (for feature consistency) and depth estimation loss (point wise L1 loss and Structural Similarity Loss).


\subsubsection{Results}
\begin{figure}[H]
    \centering
    \includegraphics[width=0.9\linewidth]{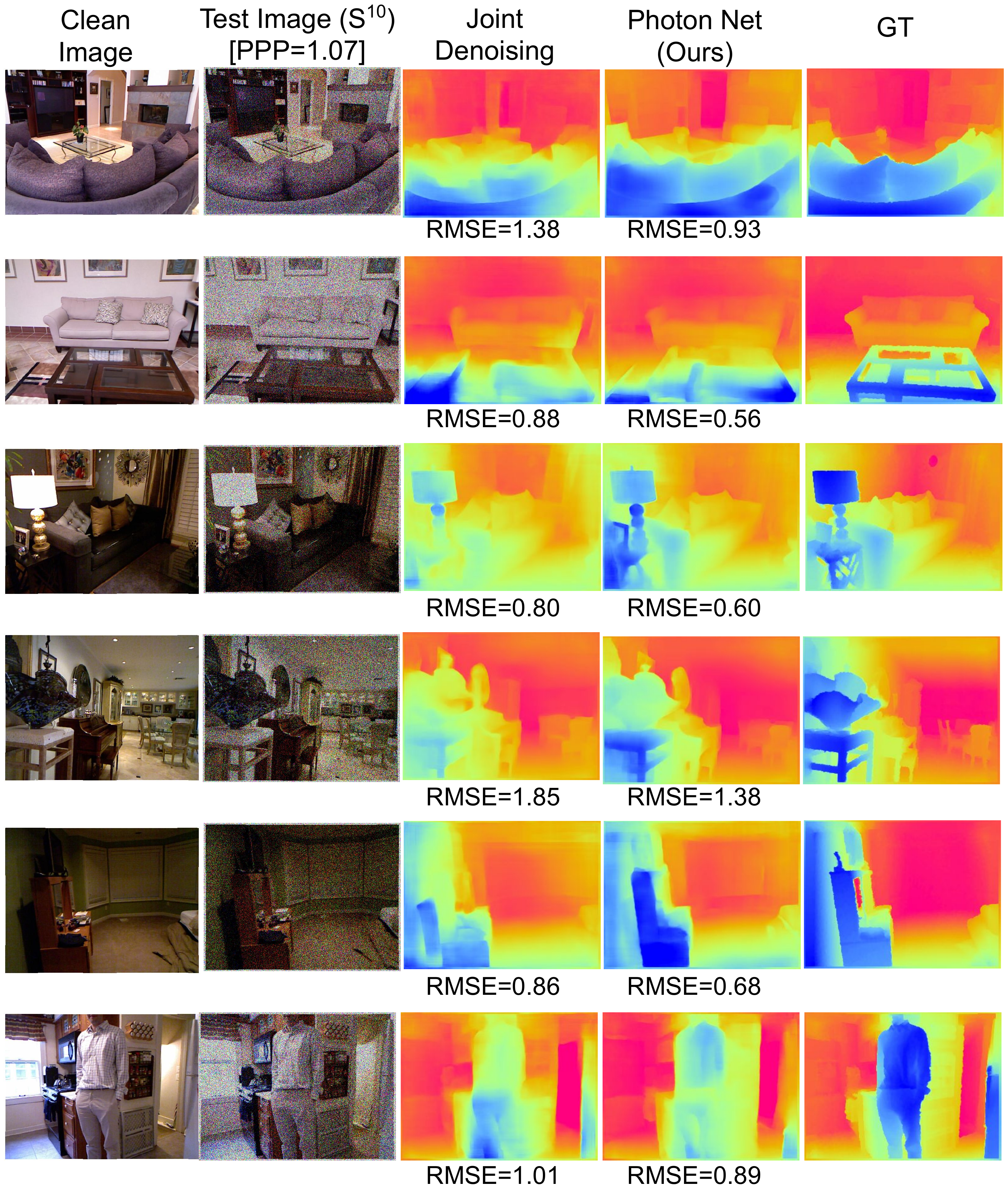}
\caption{\textbf{Monocular Depth Estimation Results} on NYUV2 dataset of $\mathcal{S}^{10}$ test images}
\label{fig:monodepthresultssupp}
\end{figure}

\begin{figure}[H]
\centering
    \includegraphics[width=\linewidth]{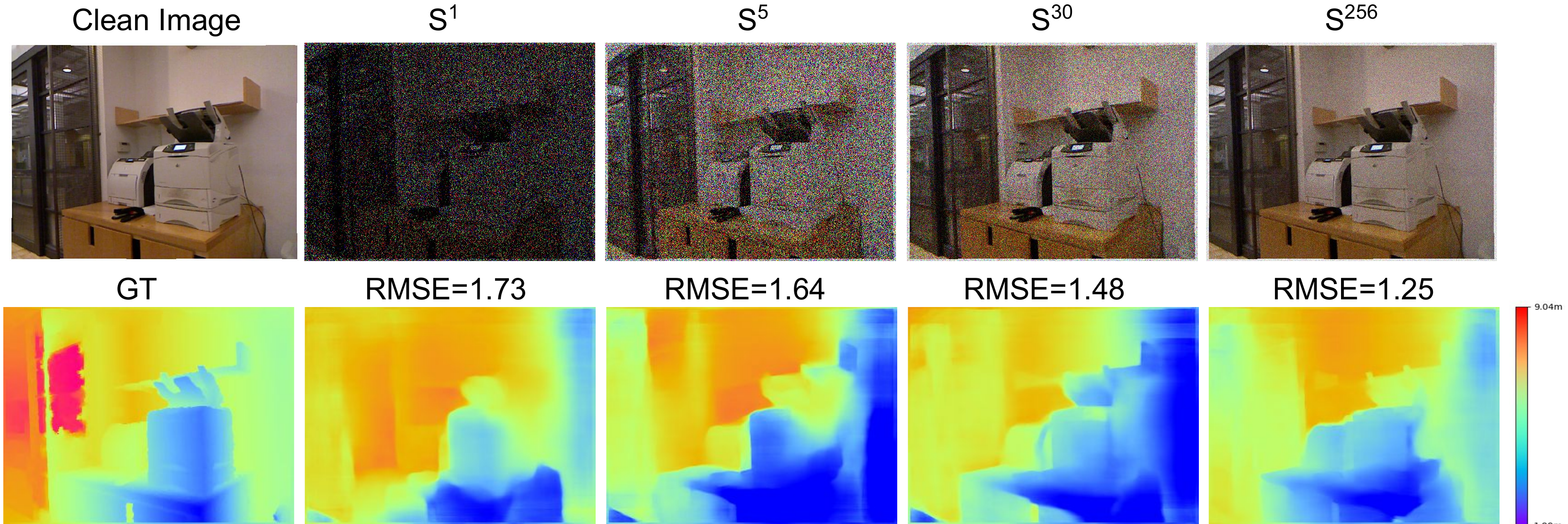}
\caption{\textbf{Monocular Depth Estimation Results} on NYUV2 dataset with increasing PPP level in the testing image}
\label{fig:monodepthresults}
\end{figure}

Figure ~\ref{fig:monodepthresultssupp} shows examples of output depth maps from the Photon Net and the baseline. Figure ~\ref{fig:monodepthresults} shows output depth maps while using higher SNR image for testing.

\subsection{Real Captures from SPADs}

\begin{figure}[H]
\centering
    \includegraphics[width=\linewidth]{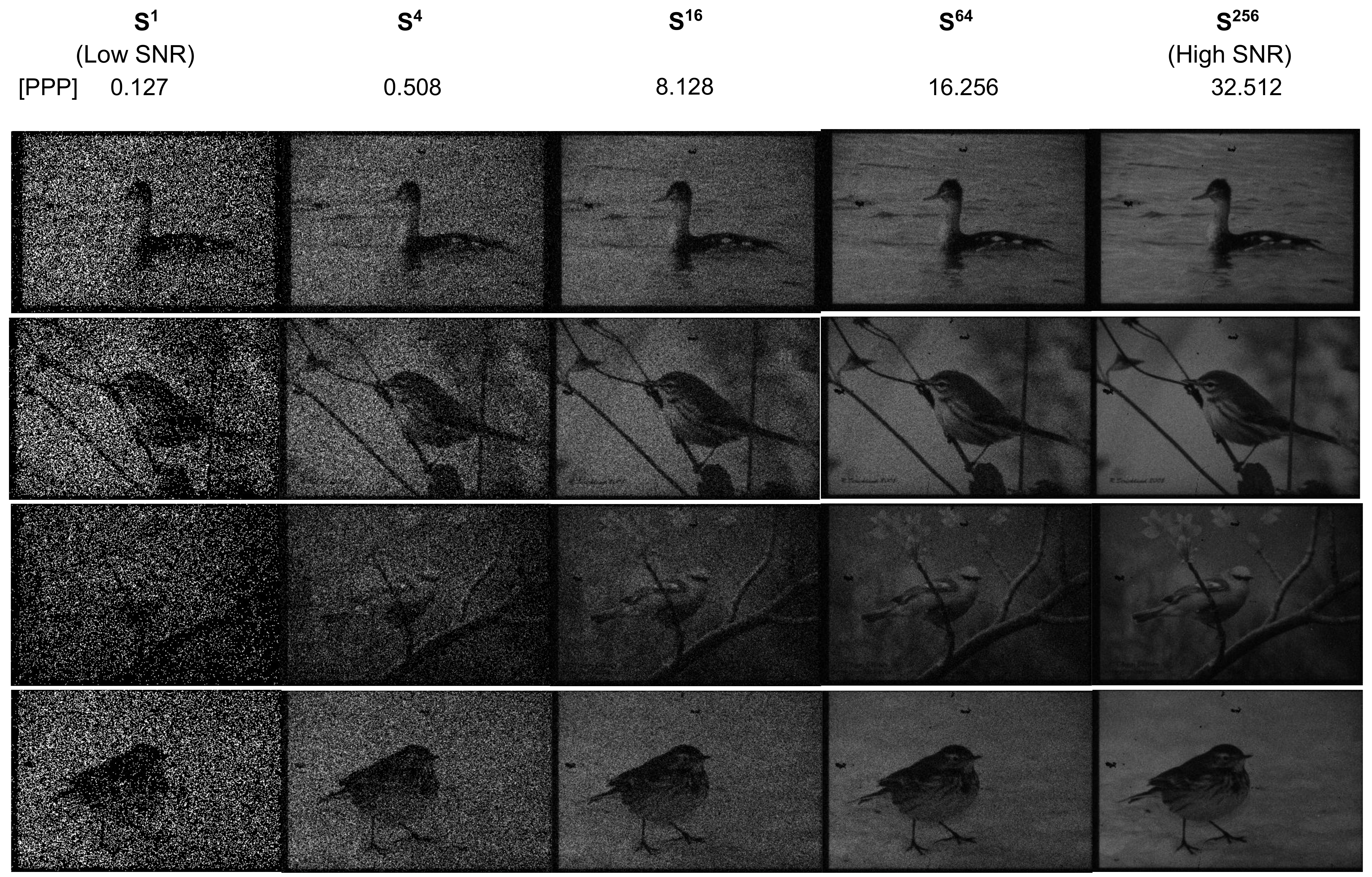}
\caption{\textbf{Real Captures:} Sample of images from SPAD cameras}
\label{fig:suppsamplespad}
\end{figure}

\begin{figure}[H]
    \centering
    \includegraphics[width=0.3\linewidth]{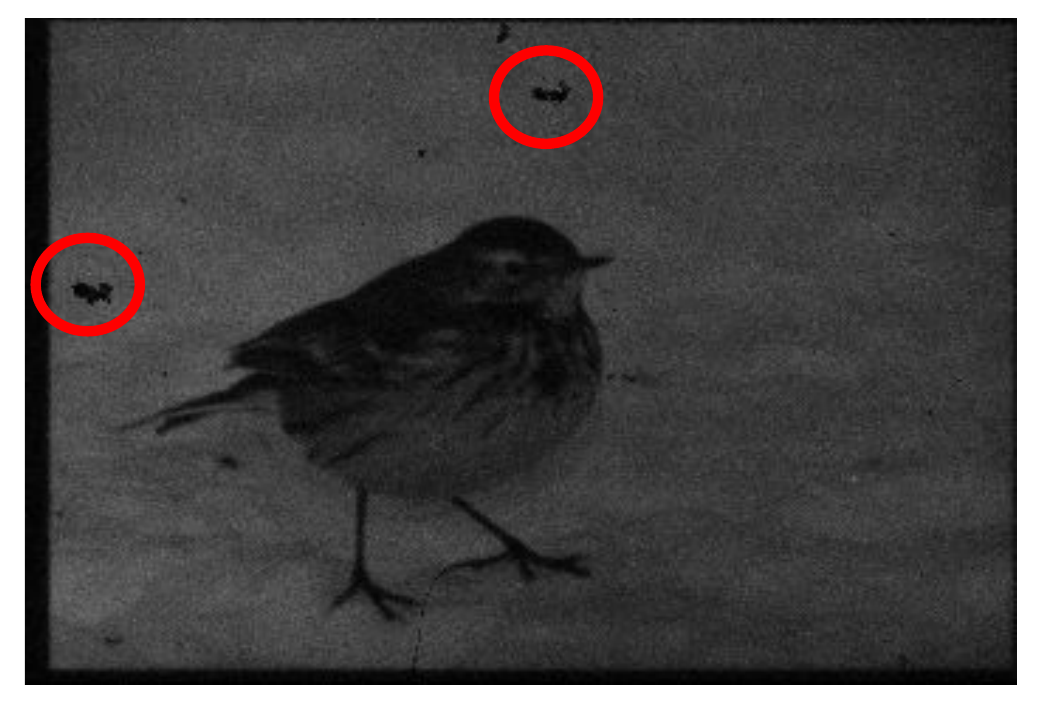}
\caption{\textbf{Artifacts} in Real Captures from SwissSPAD2 camera}
\label{fig:spadartifacts}
\end{figure}

To collect dataset of real captures from SPAD sensors, we displayed the original RGB images on a monitor screen (full screen while maintainting the aspect ratio) and captured it using SPAD sensors. The camera is positioned to cover the monitor display in its field of view. Since the monitor has the aspect ratio of 16:9 and camera has the resolution 512x256, captured frames have black padding outside the screen area. We crop all the captured frames based on the size of the original images to remove all the padding. Frames are grayscale and contain hot pixels. We correct these hot pixels by capturing an image of a black scene to identify the locations and then filter them using spatial neighborhood information.
Figure ~\ref{fig:suppsamplespad} shows example of images captured using SwissSPAD2 camera ~\cite{ulku2018512} as described in Section 7 of the main text.
Images formed from the sensor contain a few artifacts (in form of black patches) as shown in Figure ~\ref{fig:spadartifacts}.